# The Interplay of Single Ion Anisotropy and Magnetic 3$d$-4$f$ Interactions in V$^{III}_2$Ln$^{III}_2$ Butterfly Complexes

Jan Arneth,*[a]# Xian-Feng Li,[b]# Jonas Braun,*[b,c,d] Sagar Paul,[e] Michael Schulze,[e] Christopher E. Anson,[b] Wolfgang Wernsdorfer,[d,e] Annie K. Powell,*[b,c,d] Rüdiger Klingeler*[a]

# These authors have contributed equally

| | |
|---|---|
| [a] | J. Arneth, Prof. Dr. R. Klingeler |
| | Kirchhoff Institute for Physics |
| | Heidelberg University |
| | INF 227, D-69120 Heidelberg, Germany |
| | E-mail: jan.arneth@kip.uni-heidelberg.de, ruediger.klingeler@kip.uni-heidelberg.de |
| [b] | Dr. X.-F. Li, Dr. J. Braun, Dr. C. E. Anson, Prof. Dr. A. K. Powell |
| | Institute for Inorganic Chemistry |
| | Karlsruhe Institute for Technology (KIT) |
| | Kaiserstr. 12, D-76131 Karlsruhe, Germany |
| | E-mail: jonas.braun2@kit.edu, annie.powell@kit.edu |
| [c] | Dr. J. Braun, Prof. Dr. A. K. Powell |
| | Institute of Nanotechnology |
| | Karlsruhe Institute for Technology (KIT) |
| | Kaiserstr. 12, D-76131 Karlsruhe, Germany |
| | E-mail: jonas.braun2@kit.edu, annie.powell@kit.edu |
| [d] | Dr. J. Braun, Prof. Dr. A. K. Powell |
| | Institute for Quantum Materials and Technology |
| | Karlsruhe Institute for Technology (KIT) |
| | Kaiserstr. 12, D-76131 Karlsruhe, Germany |
| | E-mail: jonas.braun2@kit.edu, annie.powell@kit.edu |
| [e] | Dr. S. Paul, Dr. M. Schulze, Prof. Dr. W. Wernsdorfer |
| | Physikalisches Institut |
| | Karlsruhe Institute for Technology (KIT) |
| | Engesserstr. 15, D-76131 Karlsruhe, Germany |

**Abstract:** Within the framework of 3$d$-4$f$ molecular magnets, the most thoroughly investigated architecture is that of butterfly-shaped coordination clusters as it provides an ideal testbed to study fundamental magnetic interactions. Here, we report the synthesis and characterisation of a series of isostructural V$^{III}_2$Ln$^{III}_2$ butterfly complexes, where Ln = Y (**1$_Y$**), Tb (**2$_{Tb}$**), Dy (**3$_{Dy}$**), Ho (**4$_{Ho}$**), Er (**5$_{Er}$**), Tm (**6$_{Tm}$**), Yb (**7$_{Yb}$**), which extends the previous study on isostructural butterflies with Cr$^{III}$, Mn$^{III}$ and Fe$^{III}$. In zero external field, compounds **2$_{Tb}$**, **3$_{Dy}$** and **4$_{Ho}$** show clear maxima in the out-of-phase component of the ac susceptibility whereas small magnetic fields are needed to suppress quantum tunnelling in **6$_{Tm}$**. Combined high-field electron paramagnetic resonance spectroscopy and magnetisation measurements unambiguously reveal an easy-plane anisotropy of the V$^{III}$ ion and antiferromagnetic Ising-like 3$d$-4$f$ exchange couplings. The strength of $J_{3d-4f}$ is shown to decrease upon variation of the 4$f$ ion from Tb to Ho, while increasing antiferromagnetic interaction can be observed from Ho to Tm. The exact inverse chemical trend is found for the relative angle between the 3$d$ and 4$f$ main anisotropy axes, which highlights the important role of the lanthanide 4$f$ electron distribution anisotropy for 3$d$-4$f$ exchange.

## Introduction

One approach to enhance the performance of Single Molecule Magnets (SMMs) such as coordination clusters which exhibit magnetic hysteresis solely due to magnetisation blocking on the molecular level,[1] is the idea of designing heterometallic 3$d$-4$f$ complexes. This emerged as a promising pathway to higher blocking temperatures ($T_B$).[2-4] In these compounds the magnetic relaxation of the strongly anisotropic lanthanide moments, which is typically affected by quantum tunnelling and spin-phonon based relaxation pathways, can be slowed down via favourable magnetic interactions with the transition metal ions.[5-7] Although the combination of high effective energy barriers ($U_{eff}$) in 4$f$-based systems and slow magnetic relaxation of exchange coupled 3$d$ ion systems appears to be a logical strategy, only a few examples in which the presence of paramagnetic 3$d$ moments significantly improves the SMM properties have been reported.[8-10] Clearly the magnetic exchange coupling $J_{3d-4f}$ is one of the most important parameters influencing the performance of 3$d$-4$f$-based magnets. Hence knowledge on the underlying microscopic mechanisms is crucial for further advancements. The electronic structure of 4$f$ ions is a further complication in the determination of the size and sign of $J_{3d-4f}$ and predictions are mostly based on quantum chemical calculations.[8, 10-12] On the experimental side, high-frequency/high-field electron paramagnetic resonance (HF-EPR) spectroscopy has established itself as the main technique to investigate the magnetic 3$d$-4$f$ coupling in heterometallic complexes as it allows to quantify the coupling parameter $J_{3d\text{-}4f}$ precisely.[13-16] Other spectroscopic methods, such as inelastic neutron scattering,[17] magnetic circular dichroism[18] and far-infrared transmission spectroscopy,[19] give further insights into the microscopic parameters governing the magnetic behaviour of 3$d$-4$f$ magnets. Among the vast variety of available 3$d$-4$f$ architectures, the butterfly-motif provides a useful test-bed system and many systematic studies have been performed regarding both their static and dynamic magnetic properties.[20-24] This motif can either have the 3$d$ metal ions in the central "body" position with the 4$f$ ions at the "wingtips" (Type I) or the inverse situation for Type II. The investigation of V(III)-based butterfly systems has not been reported so far because of difficulties in obtaining stable V-Ln clusters with this desired oxidation state.

Here, we report the successful synthesis using Schlenk line techniques of a series of isostructural V$^{III}_2$Ln$^{III}_2$ Type II butterfly complexes, where Ln = Y (**1$_Y$**), Tb (**2$_{Tb}$**), Dy (**3$_{Dy}$**), Ho (**4$_{Ho}$**), Er (**5$_{Er}$**), Tm (**6$_{Tm}$**), Yb (**7$_{Yb}$**). Using combined high-field EPR and magnetisation studies we quantitatively determined the microscopic spin Hamiltonian parameters, such as 3$d$-4$f$ coupling and relative angles between the main anisotropy axes. The dependence of $J_{3d\text{-}4f}$ on the 4$f$ ion is discussed and compared to the experimental findings on other 3$d$-4$f$ complexes. We show that two distinct chemical trends can be observed for lanthanide ions with predominantly oblate and prolate anisotropy shape and conclude that the 4$f$ electron distribution plays a crucial role in the dominant 3$d$-4$f$ charge transfer process. The importance of 4$f$ charge density anisotropy becomes further evident from the magnetic relaxation behaviour as studied by ac susceptibility and low temperature µSQUID measurements.

## Experimental Detail and Methods

Single crystal crystallographic data were measured on Stoe StadiVari diffractometers equipped with Mo- or Cu-microfocus sources or a MetalJet2 liquid Ga source. Structures were solved using SHELXT[25] and refined using SHELX-2019[26] within the Olex2 platform.[27] The lattice methanols and the methyls of some ligand $t$-butyl groups were disordered and refined with sets of partial occupancy atoms. These were assigned anisotropic thermal parameters except for minor components, and refined with similarity restraints applied to bond lengths and rigid-bond restraints applied to the temperature factors. Crystallographic data are given in Table S1. Full crystallographic data and details of the structural determinations for the structures in this paper have been deposited with the Cambridge Crystallographic Data Centre as supplementary publication nos. CCDC 2406392-2406398. Copies of the data can be obtained, free of charge, from https://www.ccdc.cam.ac.uk/structures/.

The direct current (dc) and alternating current (ac) magnetisation was studied in the temperature range $T$ = 1.8 − 300 K by means of a Magnetic Properties Measurement System (MPMS3, Quantum Design) and a Physical Properties Measurement System (PPMS, Quantum Design). For all measurements powder samples were pelletised in polycarbonate capsules to avoid reorientation in external magnetic fields. The experimental data were corrected for the diamagnetic contribution of the sample holder and of the ligands calculated by means of Pascal's constants.[28] Simulations of the magnetic data were performed using the PHI software package.[29]

$M(B)$ measurements over a temperature range of 0.03–5.0 K, on single crystals of compound **3$_{Dy}$**, were conducted using µSQUIDs. A single crystal of about 100 µm (long arm) length was placed near an array of µSQUIDs on a chip, within few µm gap between the crystal edge and the µSQUID loop to ensure optimal coupling. The crystal was thermalized with Apiezon™ grease and cooled to a base temperature of 30 mK in a dilution refrigerator. A 3D vector magnet allows for the application of the magnetic field in different directions within the SQUID plane with an angular accuracy better than 0.1°. Low-temperature $M(B)$ measurements were carried out at varying field sweep rates (0.001 to 0.128 T s$^{−1}$) with a time resolution of ~1 ms. The easy axis of the crystal was identified using the 'transverse field method'.[30-31]

High-frequency/high-field electron paramagnetic (HF-EPR) resonance studies were performed using a phase sensitive millimetre vector network analyser (MVNA) by *ABmm* as a simultaneous microwave source and detector.[32] Temperature control from 2 K to 300 K was ensured by placing the sample space in the Variable Temperature Insert (VTI) of an Oxford magnet system equipped with a 16 T superconducting coil.[33] Polycrystalline powder samples were prepared in a brass ring sealed with kapton tape either as loose powder or fixed by mixing with eicosane. The former setup allows alignment of the crystallites with the external magnetic field, hence providing simplified pseudo-single-crystal spectra.[16, 34-37] Alignment is ensured by sweeping up the magnetic field to 16 T prior to each measurement and confirmed by observation of corresponding orientation jumps in the transmitted microwave intensity signal. Spectral simulations of the HF-EPR data were performed using the EasySpin software package.[38]

## Experimental Results

### Structural Analysis

Compounds **1$_Y$** to **7$_{Yb}$** with the formula [V$_2$Ln$_2$(µ$_2$-OMe)$_2$($^t$Budea)$_2$(piv)$_6$]·2CH$_3$OH (Ln$^{III}$ = Y$^{III}$, Tb$^{III}$, Dy$^{III}$, Ho$^{III}$, Er$^{III}$, Tm$^{III}$ and Yb$^{III}$) all crystallise isomorphously and thus isostructurally in the triclinic space group P$\bar{1}$ with Z = 1 (Table S1); the molecular structure of **3$_{Dy}$** is shown in Figure 1. The centrosymmetric V$_2$Dy$_2$ core corresponds to a Type II butterfly structure[24] with Dy(1) and Dy(1') occupying the body position and V(1) and V(1') are located at the wingtips. The two VDy$_2$ triangles are each bridged by a triply-bridging methoxide ligand, one above and the other below the V$_2$Dy$_2$ plane. The vanadiums are both chelated by a doubly-deprotonated ($^t$Budea)$^{2-}$ ligand with the two oxygens each bridging to one of the Dy$^{III}$ ions. The coordination is completed by four *syn,syn*-pivalates bridging the V-Dy edges of the core with two further pivalates each chelating a Dy$^{III}$ ion.

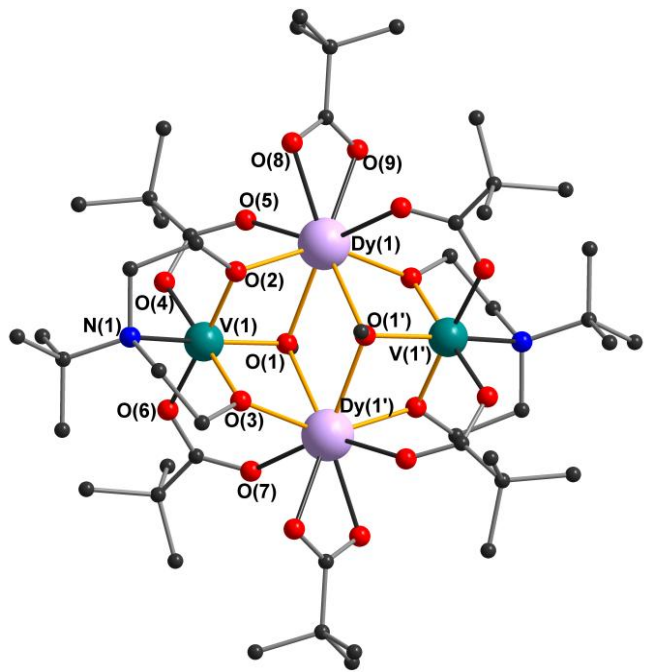

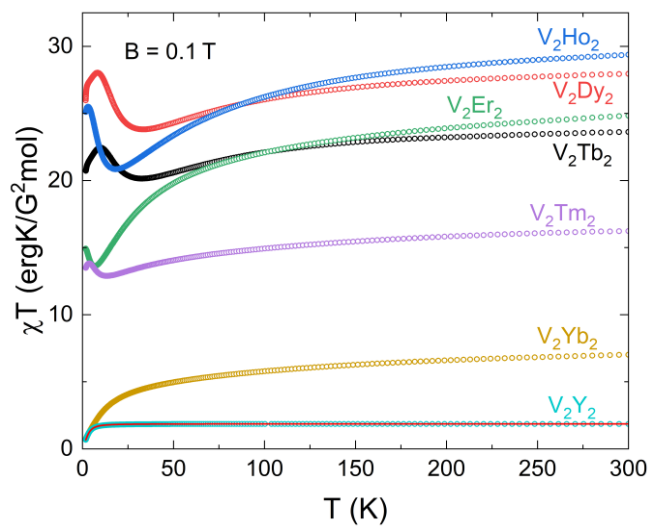

**Figure 2.** Temperature dependence of the $\chi T$-product of $1_Y$-$7_{Yb}$ in an external magnetic field of $B$ = 0.1 T. The red solid line depicts a simulation of $1_Y$ using the parameters obtained from HF-EPR (see text).

**Figure 1.** Molecular structure of the $V_2Dy_2$ cluster in $3_{Dy}$. Hydrogen atoms and minor disorder are omitted for clarity. Primed atoms at 1-$x$, 1-$y$, 1-$z$.

The clusters are isostructural to $Cr_2Dy_2$ butterflies reported by Murray *et al.*.[8] Selected bond lengths and angles are compared in Table S2, where the expected trends in the Ln-O distances given the change in size of the $Ln^{III}$ ions is clear. The $V^{III}$ oxidation states in the complexes were confirmed by Bond Valence Sum Analysis,[39] with the calculated valences for V(1) being in the range 2.94-2.97 (see Table S3).

### Direct Current (dc) Magnetisation

Variable-temperature static magnetic susceptibility measurements performed at $B$ = 0.1 T on polycrystalline powders of $1_Y$-$7_{Yb}$ result in the $\chi T$-curves shown in Figure 2. For all the measured compounds the room temperature (300 K) $\chi T$-values which are summarized in Table 1 are slightly smaller than the expected free ion values for an uncoupled system of two $V^{III}$ ions with $g$ = 2 and two $Ln^{III}$ ions. Such an observation is not surprising, since the LS-coupling in $V^{III}$ typically leads to $g$-factors smaller than 2.[40-45] In the case of $1_Y$, the magnetic data imply an average $g_{ave}$ = 1.93, which is in excellent agreement with the results of our HF-EPR measurements, as discussed later. However, the influence of spin-orbit coupling in $V^{III}$ can only partly explain the lower experimentally determined room temperature $\chi T$ of $2_{Tb}$ and $3_{Dy}$. In these compounds, the too low value of $\chi T$ might be the result of solvent molecules remaining in the crystal matrix or from the strongly anisotropic nature of the Ln moments such that even at 300 K not all $m_J$ sub-levels are equally populated.[46]

The $\chi T$-product of $1_Y$ is almost constant at temperatures above 10 K and rapidly decreases upon cooling below this value. As will be revealed by the HF-EPR studies, this drop is mainly attributed to the depopulation of the $m_J$ sub-levels, in addition to a weak antiferromagnetic interaction between the $V^{III}$ ions. A similar temperature dependence of $\chi T$ is found in $7_{Yb}$, albeit a weak decrease already begins close to room temperature and the steep drop occurs already at around 25 K. Both signatures presumably arise from the depopulation of the strongly split $m_J$ sub-levels in the anisotropic $Yb^{III}$ ions. A priori, antiferromagnetic intramolecular Yb-Yb coupling or intermolecular interaction might also yield such behaviour. However, for both we estimate an upper limit of dipolar interaction energy of less than 0.3 K which is way too small to account for the observed temperature dependence.

In contrast, the $\chi T(T)$-profiles for $2_{Tb}$-$6_{Tm}$ all exhibit a maximum at temperatures in the range 2-10 K, superposed upon the general decrease on cooling. Such behaviour could be attributed to intramolecular ferromagnetic interactions between the $V^{III}$ and the $Ln^{III}$ moments. However, the HF-EPR measurements reveal that it can actually be attributed to a parallel arrangement of the $Ln^{III}$ spins on opposite sides of the butterfly body mediated by the antiferromagnetic $V^{III}$-$Ln^{III}$ interactions. The position of the maximum in $\chi T$ gives information about the $J_{V-Ln}$ coupling strength, as discussed below.

**Table 1.** Experimentally determined and theoretical free ion $\chi T$-values (in ergK/G$^2$mol) of **1$_Y$** - **7$_{Yb}$**.

| | $g_{Ln}$ | $J_{Ln}$ | $\chi T_{exp}$ (1.8K) | $\chi T_{exp}$ (300K) | $\chi T_{\text{free ion}}$ |
|---|---|---|---|---|---|
| **1$_Y$** | | | 0.68 | 1.86 | 2.0 |
| **2$_{Tb}$** | 3/2 | 6 | 20.7 | 23.6 | 25.6 |
| **3$_{Dy}$** | 4/3 | 15/2 | 26.0 | 28.0 | 30.3 |
| **4$_{Ho}$** | 5/4 | 8 | 25.1 | 29.4 | 30.1 |
| **5$_{Er}$** | 6/5 | 15/2 | 14.9 | 24.8 | 25.0 |
| **6$_{Tm}$** | 7/6 | 6 | 13.5 | 16.2 | 16.3 |
| **7$_{Yb}$** | 8/7 | 7/2 | 0.76 | 7.0 | 7.14 |

**High-Field EPR Studies**

HF-EPR measurements on freshly ground powder samples from the same batch were carried out to determine the microscopic magnetic parameters. Figure 3 shows representative HF-EPR spectra of **1$_Y$** at variable temperature and fixed microwave frequency of $f$ = 395.4 GHz (a) as well as at variable frequency and fixed temperature of $T$ = 2 K (b). The spectra exhibit clear resonances in the accessible field and frequency range. In total two asymmetric resonance features can be identified as indicated by the vertical dashed lines in Figure 3(a). Upon increasing temperature both features become less pronounced, indicating that the initial state of the corresponding transition is (at least close to) the magnetic ground state. Moreover, the corresponding frequency dependence suggests that the low-field feature is associated with a symmetry-forbidden transition yielding $\Delta m$ = 2, while the high-field feature is related to an allowed transition with $\Delta m$ = 1. The allowed high-field feature exhibits its main spectral weight on its high-field side, thereby directly implying an easy-axis-type $g$-factor anisotropy.[47-48]

In order to quantify the qualitatively discussed parameters we performed spectral simulations using an interacting dimer Hamiltonian in the form of

$$\hat{\mathcal{H}}_{1_Y} = \sum_{i=1}^{2}\left(\mu_B \mathbf{B}\vec{g}\hat{\mathbf{S}}_i + D\hat{S}_{i,z}^2 + E(\hat{S}_{i,x}^2 - \hat{S}_{i,y}^2)\right) - J\hat{\mathbf{S}}_1\hat{\mathbf{S}}_2 \quad (1)$$

where $S_1 = S_2 = 1$, $\vec{g} = (g_\perp, g_\perp, g_\parallel)$ denotes an axial $g$-tensor, $D$ and $E$ are the axial and rhombic anisotropy constants and $J$ represents an isotropic Heisenberg interaction between the V$^{III}$ moments. Note here that both the $g$-tensor and the crystal field parameters are fixed to be equal for both V$^{III}$ sites, since their local ligand coordination, and hence the surrounding electrostatic potentials, are equivalent under point inversion. The best simulations of the experimental data were achieved by the parameters $g_\perp$ = 1.91(2), $g_\parallel$ = 1.98(2), $D$ = +9.3(4) K (6.45 cm$^{-1}$), $|E|$ = 0.25(5) K (0.17 cm$^{-1}$), $J$ = −0.18(6) K (0.12 cm$^{-1}$) and are shown as red solid lines in Figure 3. The spectral simulations confirm the easy-axis-character of the $g$-tensor, while simultaneously revealing an easy-plane single-ion anisotropy with a small rhombic distortion. The magnitude of the $D$ and $E$ parameters is not uncommon for V$^{III}$ compounds reported in the literature.[40-45] In addition, this HF-EPR analysis reveals a weak antiferromagnetic coupling between the V$^{III}$ moments which likely arises from a superexchange interaction via the diffuse empty orbitals of the diamagnetic Y$^{III}$ ion.[49] Although $D$ is the dominant term it is necessary to include the contribution of the parameters $E$ and $J$ in order to simulate the experimental data well. The parameter set extracted from HF-EPR also fully reproduces the observed magnetic data of **1$_Y$**, as can be seen by the simulation depicted in Figure 2.

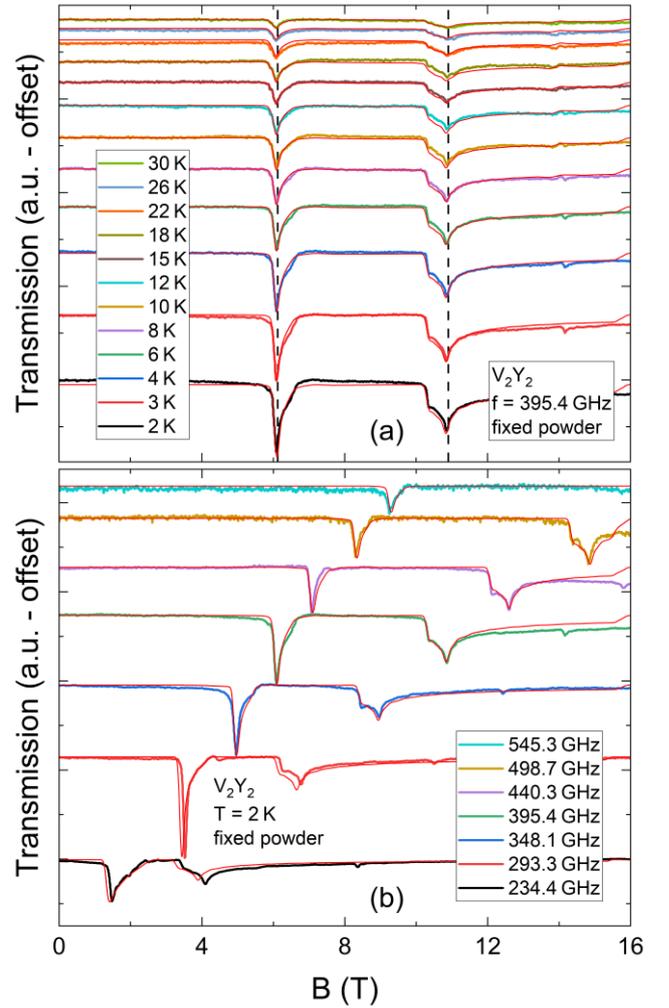

**Figure 3.** HF-EPR spectra of a fixed powder sample of **1$_Y$** at different temperatures and fixed frequency $f$ = 395.4 GHz (a) and at different frequencies and constant temperature $T$ = 2 K (b). Vertical dashed lines indicate the distinct resonance features. Red solid lines depict simulations using equation 1 with the parameters described in the main text.

We additionally performed HF-EPR spectroscopy on loose powder samples of **2$_{Tb}$-7$_{Yb}$**, for which **2$_{Tb}$**, **5$_{Er}$** and **6$_{Tm}$** show well-resolved resonance features in the accessible frequency- and field-range. The obtained resonance positions for $T$ = 2 K are summarised in the magnetic field vs. resonance frequency diagrams in Figure 4, where distinct resonance features are marked by different symbols. Compared to the spectra of $1_Y$, the resonance features in all three compounds are significantly less pronounced and exhibit strongly frequency-dependent line shapes. Upon following the main spectral weight clear resonance branches can be identified.

In $2_{Tb}$ a single EPR-transition can be observed at frequencies ranging from 280 to 500 GHz. Up to at least 8 T its resonance frequency decreases linearly with magnetic field, corresponding to an effective $g$-factor of $g_{eff} \simeq 2$, close to the $g$-value found for the $V^{III}$ ions in $1_Y$. This suggests that the transition detected here is likely to be associated with an excitation involving the $V^{III}$ ions in $2_{Tb}$. However, the data also indicate a significant magnetic coupling to the $Tb^{III}$ moments, since the extrapolated zero-field excitation gap $\Delta \simeq 515$ GHz of the resonance branch cannot be explained by the single-ion anisotropy of the $V^{III}$ ions alone. The low temperature HF-EPR spectra of $5_{Er}$ and $6_{Tm}$ exhibit up to two ($5_{Er}$) and four ($6_{Tm}$) resonance features yielding strongly curved branches with initial slopes corresponding to $g_{eff} \neq 2$. Interestingly, both complexes share a common low-energy zero-field excitation gap at $\Delta \simeq 240$ GHz, whereas the high-energy gaps differ significantly from each other. The temperature dependence of the HF-EPR spectra indicates that all observed transitions arise from either the ground or low-lying excited states (see Supplementary Material).

As a result of their large total angular momentum and complex anisotropy shape of $Ln^{III}$ ions, simulations using full spin Hamiltonians for these ions in low-symmetry ligand fields generally exceed the manageable Hilbert and parameter space. However, in suitable coordination environments an axial charge distribution can lead to well-isolated ground state doublets, which allow for a treatment of $Ln^{III}$ moments as Ising-spins at sufficiently low temperatures.[15-16, 50-51] In the frame of this model the Hamiltonian of $2_{Tb}$-$7_{Yb}$ reads

$$\hat{\mathcal{H}}_{V_2Ln_2} = \vec{R}_x^T(\theta)\hat{\mathcal{H}}_{1_Y}\vec{R}_x(\theta) + \sum_{i=3}^{4}\left(\mu_B 2\alpha \mathbf{B}\vec{\tilde{g}}\hat{\mathbf{S}}_i - 2\alpha J_{V-Ln}(\hat{S}_{1,z}\hat{S}_{i,z} + \hat{S}_{2,z}\hat{S}_{i,z})\right) \quad (2)$$

where $S_3 = S_4 = 1/2$, $\vec{\tilde{g}} = (0, 0, g_{Ln})$, $\alpha$ is a scaling factor to account for renormalization in the pseudospin transformation, and $\vec{R}_x(\theta)$ denotes a rotation of the local $1_Y$-subsystem parameters ($\vec{g}$ and $\vec{D}$) by $\theta$ around the global x-axis.

Since the local environment around the $V^{III}$ ions is essentially identical in all compounds, it is justifiable to fix the parameters obtained from the HF-EPR studies on $1_Y$ in the simulations of $2_{Tb}$-$7_{Yb}$. In addition, it is assumed that the strongly Ising-like character of the $Ln^{III}$ moments means that in a loose powder sample the crystallites can be oriented by the applied field such that the easy-anisotropy axes coincide with that of the lanthanide moments (here chosen parallel to the z-axis) and, hence, leads to an alignment of the molecules along the z-axis. The best simulations of the resonance branches are achieved by using $2\alpha J_{V-Tb} = -34.1$ K and $\theta = 0°$ for $2_{Tb}$, $2\alpha J_{V-Er} = -8.6$ K and $\theta = 60°$ for $5_{Er}$, and $2\alpha J_{V-Tm} = -12.2$ K and $\theta = 45°$ for $6_{Tm}$, respectively. A more elaborate overview of the obtained spin Hamiltonian parameters is given in Table 2. It should be noted that neither the inclusion of magnetic dipolar coupling between the $Ln^{III}$ moments nor an additional rotation of the $V^{III}$ ion subsystem around its local y-axis significantly improved the simulations.

As can be seen in Figure 4, the experimentally observed resonance branches are well described by the spin Hamiltonian in equation 2. Only the low-energy branch at high magnetic fields in $6_{Tm}$ (green squares) cannot be reproduced by our model. The slope of this branch corresponds to an effective $g$-factor of $g_{eff} \simeq 1.1$, which is close to the Landé-factor of $Tm^{III}$. Therefore, it seems reasonable that this resonance feature is associated with a transition between different $m_J$ levels of the $Tm^{III}$ ions that are effectively neglected in the Ising-approximation used here. Since $\alpha$ and $J_{V-Ln}$ are linearly correlated (see equation 2), this analysis of the HF-EPR data does not allow for a determination of the coupling constant $J_{V-Ln}$ but only of the magnetic coupling energy $2\alpha J_{V-Ln}$. In order to decouple the two parameters, we simulated the low-temperature magnetic data using the parameters obtained from the HF-EPR studies.

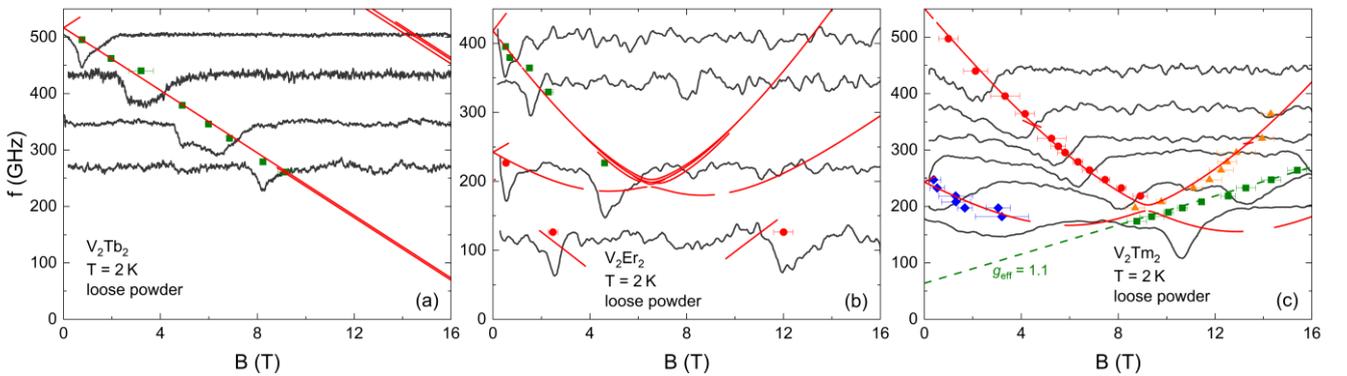

**Figure 4.** Magnetic field vs. resonance frequency plot of $2_{Tb}$ (a), $5_{Er}$ (b) and $6_{Tm}$ (c) obtained from HF-EPR measurements at $T = 2$ K. Red solid lines depict simulations using equation 2 with the parameters shown in Table 2. Selected HF-EPR spectra are visible in the background and are shifted along the ordinate to align with the corresponding measurement frequency. The green dashed line in (c) marks the resonance branch which is associated with a transition between different $m_J$ levels of the $Tm^{III}$ ions as described in the text.

Figure 5 shows the isothermal magnetisation of $2_{Tb}$ at $T$ = 1.8 K and the corresponding simulations with $\alpha$ = 6 and, hence, $J_{V\text{-}Tb}$ = −2.84 K. The chosen model is in line with the experimental data up to at least 6 T and reproduces the key characteristics of the $M(B)$-profile *i.e.* a steep increase at small magnetic fields and an inflection point, indicating a change in the magnetic ground state, at $B \simeq 5$ T. The discrepancy at higher magnetic fields is likely associated with finite contributions from the transverse components of the Ln$^{III}$ moments, *i.e.* their deviation from perfect Ising-spins in the fixed powder measurements. This can be modelled by introducing transversal components of the effective lanthanide g-tensor, such that $\vec{g} = (\tilde{g}_\perp, \tilde{g}_\perp, g_{Ln})$. As seen in Figure 5, the experimental data are well simulated using this extended model with $\tilde{g}_\perp$ = 0.32 over the entire magnetic field range.

Whereas the low-temperature $\chi T$-product of $2_{Tb}$ (inset of Figure 5) can be fitted by the simulations using the same parameters as used for the $M(B)$ simulation, the Hamiltonian in equation 2 is not able to describe the $\chi T$-product at elevated temperatures. This is not unsurprising since the population of higher-lying $m_J$ sub-levels means that the Ln$^{III}$ moments no longer equate to Ising-spins. However, the qualitative behaviour of $\chi T(T)$, especially the position of the characteristic maximum at low and minimum at intermediate temperatures, is in line with this model.

An analogous analysis of combined HF-EPR and magnetisation data on $5_{Er}$ and $6_{Tm}$ (see Fig. S2) results in the spin Hamiltonian parameters given in Table 2. In addition, despite showing no detectable HF-EPR signal, simulations of the magnetisation data also allow the determination of the microscopic parameters in $4_{Ho}$, albeit with reduced accuracy.

**Table 2.** Spin Hamiltonian parameters obtained from combined analysis of HF-EPR and magnetisation data using equation 2.

|  | $\alpha$ | $J_{V-Ln}$ (K) | $\tilde{g}_\perp$ | $\theta$ (°) |
|---|---|---|---|---|
| $2_{Tb}$ | 6 | -2.84(4) | 0.32 | 0(4) |
| $4_{Ho}$ | 7 | -0.55(11) | 0.1 | 85(16) |
| $5_{Er}$ | 6 | -0.72(7) | 0.32 | 60(8) |
| $6_{Tm}$ | 6 | -1.02(6) | 0.15 | 45(5) |

### Alternating Current (ac) Susceptibility

In order to probe their magnetic relaxation dynamics, ac susceptibility measurements with an oscillating field of $H_{ac}$ = 3 Oe were performed on all investigated complexes. Compounds $2_{Tb}$, $3_{Dy}$ and $4_{Ho}$ exhibit detectable out-of-phase ($\chi''$) signals without applied dc field. Field-induced slow relaxation of the magnetisation with an applied dc field of 0.1 T could be observed for $6_{Tm}$, suggesting that quantum tunnelling is the dominant relaxation mechanism when no external field is present. For all other compounds no $\chi''$ signals could be observed in the accessible field and frequency range. The measured frequency dependence of the out-of-phase susceptibilities are shown in Figure 6. Apart from the measurement of $3_{Dy}$ two distinct relaxation processes can clearly be discerned for all other compounds.

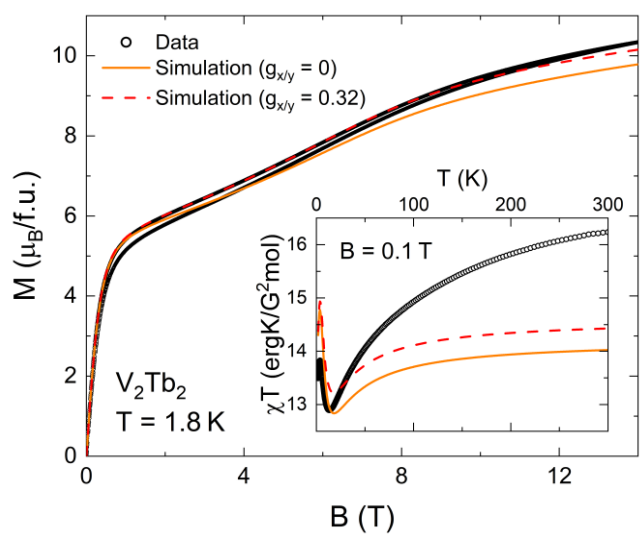

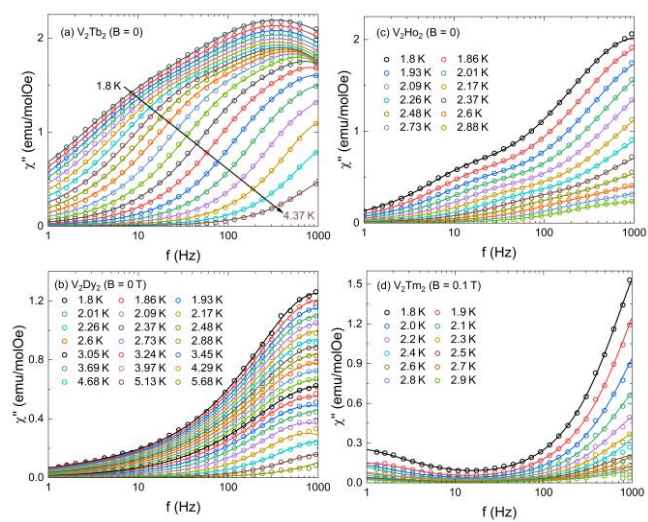

**Figure 5.** Isothermal magnetisation loop (0 T → 7 T → 0 T) of $2_{Tb}$ at $T$ = 1.8 K and simulations of the down-sweep using the strict Ising model (orange solid line) in equation 2 and an XXZ model (red dashed line) with the parameters shown in Table 2 (see the text). The inset depicts the measured and simulated $\chi T(T)$ at B = 0.1 T.

**Figure 6.** Out-of-phase ac susceptibility of $2_{Tb}$ (a), $3_{Dy}$ (b), $4_{Ho}$ (c) and $6_{Tm}$ (d) at selected temperatures. Solid lines depict fits of equation 3 to the data as described in the text. Corresponding in-phase ac susceptibility data and Cole-Cole plots are shown in the Supplementary Material.

To corroborate this assignment, the data were fitted using a generalized Double-Debye-model, *i.e.* the sum of two Debye-relaxation functions

$$\chi_{ac}(f) = \sum_{n=1}^{2}\left(\chi_{S,n} + \frac{\chi_{T,n}-\chi_{S,n}}{1+(if\tau_n)^{1-\alpha_n}}\right) \quad (3)$$

where $\chi_S$ and $\chi_T$ are the adiabatic and isothermal susceptibilities, $\tau$ denotes the relaxation time, and $0 < \alpha < 1$ qualitatively represents the distribution of relaxation times in the molecule. As can be seen from Figure 6 the chosen model describes the experimental results well. We note that one Debye function is not sufficient to reproduce the ac susceptibility of $3_{Dy}$. This implies the presence of at least two relaxation processes with similar relaxation times. The nature of the underlying microscopic mechanisms leading to slow magnetic relaxation can be elucidated by the temperature dependence of $\tau$. Qualitatively, the fast relaxation processes in $2_{Tb}$, $4_{Ho}$ and $6_{Tm}$ shift to higher frequencies with increasing temperature, while the broad maximum in $\chi''(f)$ of $3_{Dy}$ is almost temperature independent, indicative of dominant quantum tunnelling.

The relevance of under-barrier relaxation in $3_{Dy}$ can also be seen from the µSQUID measurements depicted in Fig. 7. At sub-Kelvin temperatures the single crystal $M(B)$ loops show a butterfly-shaped hysteresis which indicates fast zero field QTM and a slow thermal relaxation process in finite external magnetic fields. In agreement with our ac susceptibility data, even the fastest field scan rate of $\partial B/\partial t = 128$ mT/s yields only a small remanent magnetisation (Fig. 7(a)). However, the magnetisation loops remain open (Fig. 7(a) inset) with a small coercive field for all sweep rates. SMM behaviour in $3_{Dy}$ at such low temperatures probably arises as a result of small effective internal fields originating from intramolecular magnetic interactions. Upon heating from the lowest temperatures (Fig. 7(b)), the hysteresis region initially shrinks as it is expected when the contribution of temperature dependent relaxation processes increases. Interestingly, the inverse effect, i.e., widening of the hysteresis on increasing the temperature, can be observed in the temperature range 0.2 K < $T$ < 0.9 K. This is not a rare observation in low temperature $M(B)$ measurements when QTM and thermal relaxation times compete at an avoided level crossing region.[52-53] While the overall shape of the $M(B)$ loops at 30 mK is consistent with a predominantly ferromagnetic character, a magnetisation plateau appears near zero field as the bath temperature is increased (cf. $M(B,T=1.5$ K$)$ in the inset of Fig. 7(b)). Using the exchange field approach,[53] an antiferromagnetic Dy-Dy coupling of $J_{Dy-Dy}$ = -6(1) mK can be read off the data (for a more detailed analysis and discussion see the SI).

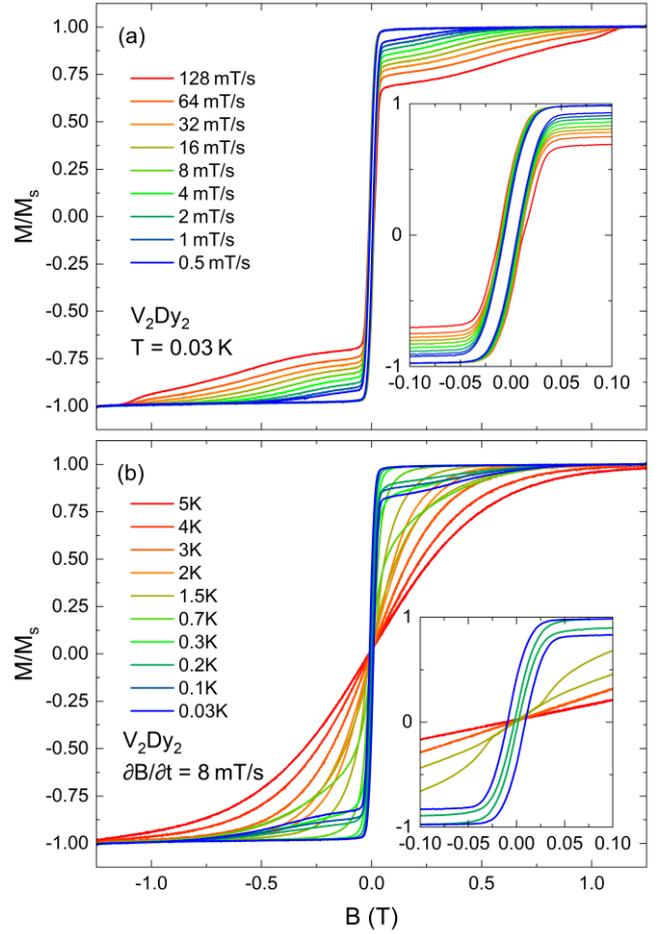

**Figure 7.** $M(B)$ measurements on a single crystal of $3_{Dy}$, at fixed bath temperature $T$ = 0.03 K and different sweep rates of magnetic field (a), and fixed sweep rate $\partial B/\partial t$ = 8 mT/s and different bath temperatures (b). The insets show the enlarged version of corresponding figures considering only a few curves for clarity.

A more quantitative investigation of the magnetic relaxation can be provided for both relaxation processes in $2_{Tb}$ and for the slow process in $4_{Ho}$ from a plot of $\tau$ vs $T$ (cf. Figure 8) using the relaxation times extracted from the Debye fits. The linearity of ln($\tau$) vs 1/$T$ at high temperatures implies the presence of Orbach and/or slow Raman relaxation. In contrast, faster relaxation mechanisms, such as quantum tunnelling or fast Raman relaxation, dominate at low temperatures. We fit our data over the whole temperature range using

$$\tau^{-1} = \tau_{QTM}^{-1} + \tau_R^{-1}\frac{e^{-W/k_BT}}{\left(1+e^{-W/k_BT}\right)^2} + \tau_O^{-1}e^{-U_O/k_BT} \quad (4)$$

Here, the first term denotes the temperature independent quantum tunnelling of magnetisation, the second term describes Raman relaxation involving only energetically low-lying optical phonons as suggested independently by Gu and Wu and by Lunghi *et al.*[54-56] and the third term represents Orbach relaxation. The resulting fit parameters, summarised in Table 3, indicate that at low temperatures quantum tunnelling is the main contribution to the fast relaxation process in $2_{Tb}$, whereas the slower process is dominated by slow Raman relaxation. This finding is further corroborated by ac susceptibility measurements under a 0.1 T dc field showing that only one relaxation process is visible, which is mainly governed by Orbach and Raman relaxation (see Supplementary Material). At elevated temperatures magnetic relaxation happens predominantly via Orbach relaxation with effective barriers of $U_{eff}$ = 14.6 K, 23.6 K (fast and slow process $2_{Tb}$), and 25.8 K ($4_{Ho}$), for the three sets of relaxation processes shown in Figure 8. These values are comparable to the barrier heights observed in other butterfly-shaped heterometallic 3d-4f complexes.[21]

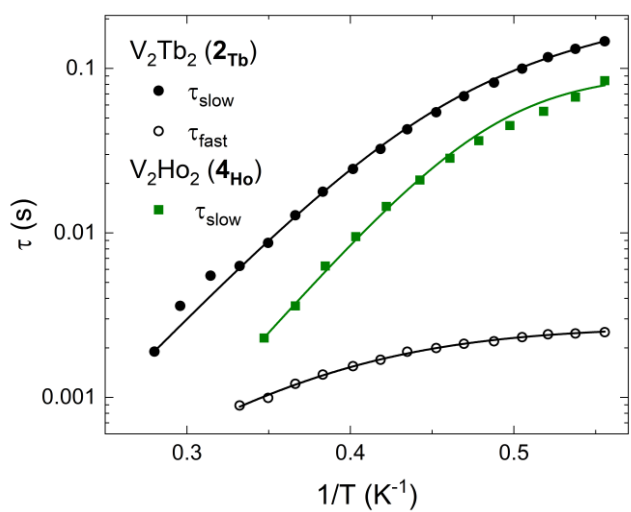

**Figure 8.** Temperature dependence of the relaxation times in $2_{Tb}$ and $4_{Ho}$ in the absence of an external static magnetic field. Solid lines depict fits to the data using equation 4 as described in the text.

The absence of a $\chi''$ signal in $1_Y$ directly indicates that slow magnetic relaxation in the $V_2Ln_2$ compounds can only be observed when paramagnetic $Ln^{III}$ ions are present. Hence, a number of conclusions can be drawn from the presented ac susceptibility data: (i) Slow magnetic relaxation in zero field occurs only in $2_{Tb}$, $3_{Dy}$ and $4_{Ho}$, implying that the ligand field around the $Ln^{III}$ ions is more suitable for oblate charge distributions. (ii) Despite being predominantly prolate, the 4f-orbitals of $Tm^{III}$ in $6_{Tm}$ experience a considerable axial crystal field anisotropy, such that only small external magnetic fields are needed to suppress quantum tunnelling. (iii) Sizable magnetic coupling to the $V^{III}$ moments is not sufficient to slow down the magnetic relaxation to the Hz or even sub-Hz scale.

**Table 3.** Parameters obtained by the fit of the temperature dependence of the relaxation times in $2_{Tb}$ and $4_{Ho}$ using equation 4.

| | $\tau_{QTM}$ (s) | $\tau_R$ (s) | $W$ (K) | $\tau_O$ (s) | $U_O$ (K) |
|---|---|---|---|---|---|
| $2_{Tb}$ - $\tau_{fast}$ | 2.7×10⁻³ | | | 1.0×10⁻⁵ | 14.6 |
| $2_{Tb}$ - $\tau_{slow}$ | | 1.06 | 0.7 | 2.6×10⁻⁶ | 23.6 |
| $4_{Ho}$ - $\tau_{slow}$ | 0.1 | | | 3.1×10⁻⁷ | 25.8 |

## Discussion

The discussion of 3d-4f coupling beyond dipole-dipole-interaction in the literature was initially based on a model proposed for a $Gd_2Cu_4$ complex by Kahn et al.[49] with similar ideas being reported by Gatteschi et al..[57-58] In summary, the 3d-4f superexchange interaction arises from two distinct exchange mechanisms. The first of these is electron transfer from the occupied 3d orbitals of the transition metal ion into the empty 5d orbitals of the lanthanide favouring a ferromagnetic spin alignment due to enhanced Hund's coupling in the charge transfer state, as long as the 4f shell is at least half-filled.[49] The second is excitation of the transition metal 3d electrons into the partially filled lanthanide 4f orbitals leading to either a ferromagnetic or an antiferromagnetic contribution to $J_{3d-4f}$ depending on whether the corresponding orbitals are orthogonal to each other or overlap. Taking into account the influence of single ion anisotropy on the 3d orbitals, this model was further developed to include also other transition metal ions with $S$ > 1/2.[6, 14-15] Considering the magnitude of $J_{V-Ln}$ obtained from HF-EPR (Table 2), it becomes evident that dipolar interactions are negligible. Hence, the observation of antiferromagnetic interaction implies that electron exchange between 3d and 4f orbitals is the dominant coupling mechanism. This finding can be rationalised by the predominant π-character of the occupied $V^{III}$ $t_{2g}$ orbitals, which renders the ferromagnetic σ-type 3d-5f charge transfer rather inefficient[59] and favours direct orbital overlap with the $Ln^{III}$ 4f orbitals.[60]

Figure 9(a) summarises the results of our HF-EPR studies by showing the dependence of $J_{V-Ln}$ and $\theta$ on the lanthanide ion. Of the studied complexes $2_{Tb}$ exhibits the largest magnetic interaction between $V^{III}$ and $Ln^{III}$ moments. This coupling becomes significantly weaker when $Tb^{III}$ is replaced by $Ho^{III}$ and subsequently increases for the heavier lanthanides $Er^{III}$ and $Tm^{III}$. Interestingly, the relative angle between the vanadium and lanthanide main anisotropy axes $\theta$ exhibits the inverse trend. According to Kahn's model[49] the ferromagnetic contribution to $J_{V-Ln}$ is expected to decrease with increasing atomic number from $Gd^{III}$ onwards as the amount of unpaired 4f electrons, and hence Hund's coupling in the charge transfer state $4f^n5d^1$, decreases along the series. While this mechanism qualitatively rationalises the increase of overall antiferromagnetic coupling from $Ho^{III}$ to $Tm^{III}$, the finding of $2_{Tb}$ exhibiting the largest $J_{V-Ln}$ represents an outlier. The reason for this is (likely) to be the decreasing orbital overlap between the partially filled 3d and 4f orbitals due to a rotation of the $Ln^{III}$ easy anisotropy axis (see Figure 9). A similar mechanism has been found to explain the dependence of $J_{Cr-Gd}$ on the Cr-F-Gd bond angle in $Cr^{III}Gd^{III}$ complexes, where the 3$d$-4$f$ orbital overlap is mainly varied by a rotation of the $Cr^{III}$ anisotropy tensor via changes in the ligand field.[59]

In the literature, HF-EPR has established itself as the main method of choice for quantitative experimental determination of 3$d$-4$f$ coupling.[14-16, 50-51, 61-62] An overview of the $J_{M-Ln}$ (M = transition metal) in different heterometallic complexes as obtained in recent HF-EPR studies is given in Table 4 and visualised in Figure 9(b). In most of the investigated compounds, namely those with $Cu^{II}$, $Ni^{II}$ and $Fe^{III}$ as the 3$d$ ion, ferromagnetic 3$d$-4$f$ interaction is found and rationalised by a dominance of the 3$d$ to 5$d$ charge transfer process.[13-16, 61] In contrast, the results on $V^{III}$-based butterfly complexes presented here evidence antiferromagnetic coupling between the 3$d$ and 4$f$ moments, which has also been observed in dinuclear $V^{IV}$-Ln molecules.[13] As mentioned above, the qualitative differences probably arise from different spatial distributions of the contributing 3$d$ orbitals, e.g., $3d_{x^2-y^2}$ for $Cu^{II}$ and $3d_{xy}$ for $V^{IV/III}$, that determine the absence or presence of overlap with the $Ln^{III}$ orbitals.[59-60] Similarly, the size of $J_{M-Ln}$ depends on the coordination geometry and on the influence of other magnetic orbitals in the corresponding electronic configuration. Whereas the expected chemical trend of decreasing $J_{M-Ln}$ for heavier lanthanide ions is generally confirmed by the experimental data, the studies are usually restricted to the series $Ln^{III}$ = $Ce^{III}$-$Ho^{III}$. Here, we expand this range to include the $Ln^{III}$ ions beyond $Gd^{III}$ with prolate electron density distribution, i.e. $Er^{III}$ and $Tm^{III}$. In fact, the two regimes observed in Figure 9 can be related to the 4$f$ anisotropy shape for the later $Ln^{III}$ ions: For the oblate Ln ions $J_{V-Ln}$ weakens with the Ln atomic number, but increases upon the series when the 4$f$ electron distribution exhibits a prolate shape. The correlation of 4$f$ anisotropy shape and magnetic 3$d$-4$f$ coupling in the $V^{III}_2Ln_2$ butterfly series can be rationalised by considering changes in the $Ln^{III}$ anisotropy axis. This leads to variation in orbital overlap as a result of ligand field effects, as described earlier.

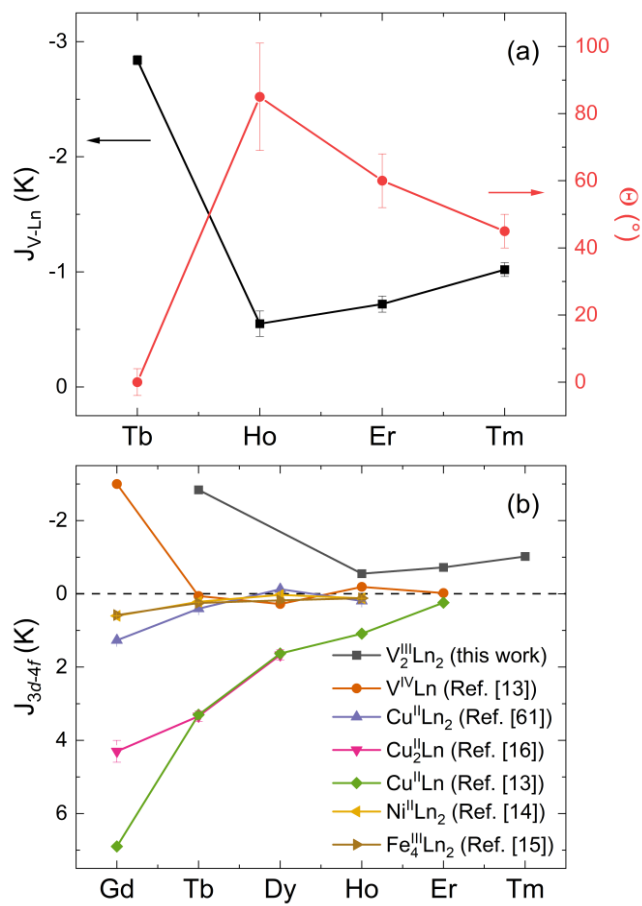

**Figure 9.** (a) Chemical trend of the $V^{III}$-$Ln^{III}$ coupling constant and of the relative angle between the main anisotropy axes (see equation 2) as determined by HF-EPR. (b) Experimentally determined $J_{3d-4f}$ in different heterometallic complexes as reported in recent HF-EPR studies.

**Table 4.** Magnetic 3$d$-4$f$ coupling in selected heterometallic molecular complexes as reported here and in the literature.

| | $J_{M-Ln}$ (K) | | | | | | |
|---|---|---|---|---|---|---|---|
| Ln = | Gd | Tb | Dy | Ho | Er | Tm | Ref. |
| $V^{III}_2Ln_2$ | | -2.84(4) | | -0.55(11) | -0.72(7) | -1.02(6) | this work |
| $V^{IV}Ln$ | -3.0 | 0.06(1) | 0.28(1) | -0.19 | -0.02(1) | | 13 |
| $Cu^{II}Ln_2$ | 1.271(7) | 0.405(3) | -0.126(3) | 0.196(13) | | | 61 |
| $Cu^{II}_2Ln$ | 4.3(3) | 3.34(14) | 1.67(14) | | | | 16 |
| $Cu^{II}Ln$ | 6.9 | >3.3 | 1.63(1) | 1.09(2) | 0.24(1) | | 13 |
| $Ni^{II}Ln_2$ | 0.602(8) | 0.216(12) | 0.031 | 0.122(3) | | | 14 |
| $Fe^{III}_4Ln_2$ | 0.58 | 0.25(5) | 0.18(8) | 0.12(8) | | | 15 |

## Conclusion

We report the successful synthesis of a series of butterfly-motif $V^{III}_2Ln^{III}_2$ ($Ln^{III}$ = $Tb^{III}$-$Yb^{III}$, $Y^{III}$) complexes and their magnetic characterisation *via* combined magnetic SQUID, µSQUID and HF-EPR studies. In particular, our analysis allows us to quantify 3$d$-4$f$ coupling by treating the $Ln^{III}$ moments as essentially Ising spins. Unlike for other 3$d$-4$f$ complexes, $J_{V-Ln}$ does not monotonically decrease upon increasing Ln atomic number but is shown to exhibit opposite effects for oblate and prolate Ln 4$f$ orbital distributions. This trend is rationalised through the variation in orbital overlap due to a rotation of the $Ln^{III}$ anisotropy axis. The important role of 4$f$ anisotropy shape is also confirmed by the ac susceptibility data which show the absence of slow magnetic relaxation in zero field for prolate $Ln^{III}$ ions.

## Supporting Information

The authors have referred to reference[53] and reference[39] and cited an additional reference within the Supporting Information.[63]

Details of the syntheses and crystallography together with Bond Valence Sum analyses and additional magnetic and HF-EPR data are provided in the Supplementary Information. Full crystallographic data and details of the structural determinations for the structures in this paper have been deposited with the Cambridge Crystallographic Data Centre as supplementary publication nos. CCDC 2406392-2406398. Copies of the data can be obtained, free of charge, from https://www.ccdc.cam.ac.uk/structures/

## Acknowledgements

Measurements were supported by Deutsche Forschungsgemeinschaft (DFG) under Germany's Excellence Strategy EXC2181/1-390900948 (the Heidelberg STRUCTURES Excellence Cluster). J. A. acknowledges support by the International Max-Planck Research School for Quantum Dynamics (IMPRS-QD) Heidelberg. X.L., J.B., C.E.A and A.K.P acknowledge funding from the DFG CRC 1573 "4f for Future" and the Helmholtz Foundation POF MSE. X.-F.L. thanks the China Scholarship Council (CSC) for a scholarship. W.W. thanks the German Research Foundation (DFG) for the Gottfried Wilhelm Leibniz-Award, ZVN-2020_WE 4458-5.

[1] R. Sessoli, D. Gatteschi, A. Caneschi, M. A. Novak, *Nature* **1993**, *365*, 141-143.
[2] T. Kido, Y. Ikuta, Y. Sunatsuki, Y. Ogawa, N. Matsumoto, N. Re, *Inorg. Chem.* **2002**, *42*, 398-408.
[3] S. Osa, T. Kido, N. Matsumoto, N. Re, A. Pochaba, J. Mrozinski, *J. Am. Chem. Soc.* **2004**, *126*, 420-421.
[4] L. R. Piquer, E. C. Sanudo, *Dalton Trans.* **2015**, *44*, 8771-8780.
[5] X. L. Li, F. Y. Min, C. Wang, S. Y. Lin, Z. Liu, J. Tang, *Inorg. Chem.* **2015**, *54*, 4337-4344.
[6] T. Gupta, M. F. Beg, G. Rajaraman, *Inorg. Chem.* **2016**, *55*, 11201-11215.
[7] S. K. Singh, M. F. Beg, G. Rajaraman, *Chem. Eur. J.* **2016**, *22*, 672-680.
[8] S. K. Langley, D. P. Wielechowski, V. Vieru, N. F. Chilton, B. Moubaraki, B. F. Abrahams, L. F. Chibotaru, K. S. Murray, *Angew. Chem. Int. Ed.* **2013**, *52*, 12014-12019.
[9] S. K. Langley, D. P. Wielechowski, V. Vieru, N. F. Chilton, B. Moubaraki, L. F. Chibotaru, K. S. Murray, *Chem. Sci.* **2014**, *5*, 3246-3256.
[10] Y. Peng, M. K. Singh, V. Mereacre, C. E. Anson, G. Rajaraman, A. K. Powell, *Chem. Sci.* **2019**, *10*, 5528-5538.
[11] J. Rinck, G. Novitchi, W. Van den Heuvel, L. Ungur, Y. Lan, W. Wernsdorfer, C. E. Anson, L. F. Chibotaru, A. K. Powell, *Angew. Chem. Int. Ed.* **2010**, *49*, 7583-7587.
[12] J. L. Liu, J. Y. Wu, Y. C. Chen, V. Mereacre, A. K. Powell, L. Ungur, L. F. Chibotaru, X. M. Chen, M. L. Tong, *Angew. Chem. Int. Ed.* **2014**, *53*, 12966-12970.
[13] R. Watanabe, K. Fujiwara, A. Okazawa, G. Tanaka, S. Yoshii, H. Nojiri, T. Ishida, *Chem. Commun.* **2011**, *47*, 2110-2112.
[14] A. Okazawa, T. Shimada, N. Kojima, S. Yoshii, H. Nojiri, T. Ishida, *Inorg. Chem.* **2013**, *52*, 13351-13355.
[15] S. F. M. Schmidt, C. Koo, V. Mereacre, J. Park, D. W. Heermann, V. Kataev, C. E. Anson, D. Prodius, G. Novitchi, R. Klingeler, A. K. Powell, *Inorg. Chem.* **2017**, *56*, 4796-4806.
[16] N. Ahmed, T. Sharma, L. Spillecke, C. Koo, K. U. Ansari, S. Tripathi, A. Caneschi, R. Klingeler, G. Rajaraman, M. Shanmugam, *Inorg. Chem.* **2022**, *61*, 5572-5587.
[17] F. J. Kettles, V. A. Milway, F. Tuna, R. Valiente, L. H. Thomas, W. Wernsdorfer, S. T. Ochsenbein, M. Murrie, *Inorg. Chem.* **2014**, *53*, 8970-8978.
[18] X. Wang, S. Q. Wang, J. N. Chen, J. H. Jia, C. Wang, K. Paillot, I. Breslavetz, L. S. Long, L. Zheng, G. Rikken, C. Train, X. J. Kong, M. Atzori, *J. Am. Chem. Soc.* **2022**, *144*, 8837-8847.
[19] T. D. Kang, E. C. Standard, P. D. Rogers, K. H. Ahn, A. A. Sirenko, A. Dubroka, C. Bernhard, S. Park, Y. J. Choi, S. W. Cheong, *Phys. Rev. B* **2012**, *86*, 144112.
[20] S. K. Langley, L. Ungur, N. F. Chilton, B. Moubaraki, L. F. Chibotaru, K. S. Murray, *Inorg. Chem.* **2014**, *53*, 4303-4315.
[21] S. K. Langley, D. P. Wielechowski, N. F. Chilton, B. Moubaraki, K. S. Murray, *Inorg. Chem.* **2015**, *54*, 10497-10503.
[22] E. Moreno Pineda, N. F. Chilton, F. Tuna, R. E. Winpenny, E. J. McInnes, *Inorg. Chem.* **2015**, *54*, 5930-4591.
[23] Y. Peng, H. Kaemmerer, A. K. Powell, *Chem. Eur. J.* **2021**, *27*, 15043-15065.
[24] Y. Peng, A. K. Powell, *Coord. Chem. Rev.* **2021**, *426*, 213490.
[25] G. M. Sheldrick, *Acta Cryst. A* **2015**, *71*, 3-8.
[26] G. M. Sheldrick, *Acta Crystallogr. C* **2015**, *71*, 3-8.
[27] O. V. Dolomanov, L. J. Bourhis, R. J. Gildea, J. A. K. Howard, H. Puschmann, *J. Appl. Cryst.* **2009**, *42*, 339-341.
[28] G. A. Bain, J. F. Berry, *J. Chem. Educ.* **2008**, *85*, 532-536.
[29] N. F. Chilton, R. P. Anderson, L. D. Turner, A. Soncini, K. S. Murray, *J. Comput. Chem.* **2013**, *34*, 1164-1175.
[30] W. Wernsdorfer, N. E. Chakov, G. Christou, *Phys. Rev. B* **2004**, *70*, 132413.
[31] W. Wernsdorfer, *Superconductor Science and Technology* **2009**, *22*, 064013.
[32] P. Comba, M. Grosshauser, R. Klingeler, C. Koo, Y. Lan, D. Muller, J. Park, A. Powell, M. J. Riley, H. Wadepohl, *Inorg. Chem.* **2015**, *54*, 11247-11258.
[33] J. Werner, W. Hergett, M. Gertig, J. Park, C. Koo, R. Klingeler, *Phys. Rev. B* **2017**, *95*, 214414.
[34] A. L. Barra, P. Debrunner, D. Gatteschi, C. E. Schulz, R. Sessoli, *Europhys. Lett.* **1996**, *35*, 133-138.
[35] D. P. Goldberg, J. Telser, J. Kryzstek, A. G. Montalban, L.-C. Brunel, A. G. M. Barett, B. M. Hoffman, *J. Am. Chem. Soc.* **1997**, *119*, 8722-8723.

[36] L. Spillecke, C. Koo, O. Maximova, V. S. Mironov, V. A. Kopotkov, D. V. Korchagin, A. N. Vasiliev, E. B. Yagubskii, R. Klingeler, *Dalton Trans.* **2021**, *50*, 18143-18154.

[37] L. Spillecke, S. Tripathi, C. Koo, M. Ansari, S. Vaidya, A. Rasamsetty, T. Mallah, G. Rajaraman, M. Shanmugam, R. Klingeler, *Polyhedron* **2021**, *208*, 115389.

[38] S. Stoll, A. Schweiger, *J. Magn. Reson.* **2006**, *178*, 42-55.

[39] W. Liu, H. H. Thorp, *Inorg. Chem.* **1993**, *32*, 4102-4105.

[40] J. Krzystek, A. T. Fiedler, J. J. Sokol, A. Ozarowski, S. A. Zvyagin, T. C. Brunold, J. R. Long, L.-C. Brunel, J. Telser, *Inorg. Chem.* **2004**, *43*, 5654-5658.

[41] R. Beaulac, P. L. W. Tregenna-Piggott, A.-L. Barra, H. Weihe, D. Luneau, C. Reber, *Inorg. Chem.* **2006**, *45*, 3399-3407.

[42] J. Krzystek, A. Ozarowski, J. Telser, D. C. Crans, *Coord. Chem. Rev.* **2015**, *301-302*, 123-133.

[43] T. A. Bazhenova, L. V. Zorina, S. V. Simonov, V. S. Mironov, O. V. Maximova, L. Spillecke, C. Koo, R. Klingeler, Y. V. Manakin, A. N. Vasiliev, E. B. Yagubskii, *Dalton Trans.* **2020**, *49*, 15287-15298.

[44] M. R. Saber, K. Thirunavukkuarasu, S. M. Greer, S. Hill, K. R. Dunbar, *Inorg. Chem.* **2020**, *59*, 13262-13269.

[45] Z. Janas, J. Jezierska, A. Ozarowski, A. Bienko, T. Lis, A. Jezierski, M. Krawczyk, *Dalton Trans.* **2021**, *50*, 5184-5196.

[46] J. Tang, P. Zhang, *Lanthanide Single Molecule Magnets*, Springer Verlag, Berlin Heidelberg, **2015**.

[47] A. Abragam, B. Bleaney, *Electron Paramagnetic Resonance of Transition Ions*, Oxford University Press, Oxford, **2012**.

[48] P. Bertrand, *Electron Paramagnetic Resonance Spectroscopy - Fundamentals*, Springer Nature, Cham, **2020**.

[49] M. Andruh, I. Ramade, E. Codjovi, O. Guillou, O. Kahn, J. C. Trombe, *J. Am. Chem. Soc.* **1993**, *115*, 1822-1829.

[50] R. T. Galeev, L. V. Mingalieva, A. A. Sukhanov, V. K. Voronkova, Y. Peng, A. K. Powell, *Appl. Magn. Reson.* **2019**, *50*, 1429-1441.

[51] P. Comba, M. Enders, M. Grosshauser, M. Hiller, R. Klingeler, C. Koo, D. Muller, G. Rajaraman, A. Swain, M. Tavhelidse, H. Wadepohl, *Chem. Eur. J.* **2021**, *27*, 9372-9382.

[52] E. Moreno-Pineda, G. Taran, W. Wernsdorfer, M. Ruben, *Chem Sci* **2019**, *10*, 5138-5145.

[53] Z. Zhu, S. Paul, C. Zhao, J. Wu, X. Ying, L. Ungur, W. Wernsdorfer, F. Meyer, J. Tang, *J. Am. Chem. Soc.* **2024**, *146*, 18899-18904.

[54] L. Gu, R. Wu, *Phys. Rev. Lett.* **2020**, *125*, 117203.

[55] M. Briganti, F. Santanni, L. Tesi, F. Totti, R. Sessoli, A. Lunghi, *J. Am. Chem. Soc.* **2021**, *143*, 13633-13645.

[56] L. Gu, R. Wu, *Phys. Rev. B* **2021**, *103*, 014401.

[57] A. Bencini, C. Benelli, A. Caneschi, R. L. Carlin, A. Dei, D. Gatteschi, *J. Am. Chem. Soc.* **1985**, *107*, 8128-8136.

[58] G. Rajaraman, F. Totti, A. Bencini, A. Caneschi, R. Sessoli, D. Gatteschi, *Dalton Trans.* **2009**, 3153-3161.

[59] S. K. Singh, K. S. Pedersen, M. Sigrist, C. A. Thuesen, M. Schau-Magnussen, H. Mutka, S. Piligkos, H. Weihe, G. Rajaraman, J. Bendix, *Chem. Commun.* **2013**, *49*, 5583-5585.

[60] S. K. Singh, G. Rajaraman, *Dalton Trans.* **2013**, *42*, 3623-3630.

[61] A. Okazawa, R. Watanabe, M. Nezu, T. Shimada, S. Yoshii, H. Nojiri, T. Ishida, *Chem. Lett.* **2010**, *39*, 1331-1332.

[62] T. Ishida, R. Watanabe, K. Fujiwara, A. Okazawa, N. Kojima, G. Tanaka, S. Yoshii, H. Nojiri, *Dalton Trans.* **2012**, *41*, 13609-13619.

[63] C. F. Macrae, I. Sovago, S. J. Cottrell, P. T. A. Galek, P. McCabe, E. Pidcock, M. Platings, G. P. Shields, J. S. Stevens, M. Towler, P. A. Wood, *J. Appl. Crystallogr.* **2020**, *53*, 226-233

# Supplemental Information: The Interplay of Single Ion Anisotropy and Magnetic 3d-4f Interactions in V$^{III}_2$Ln$^{III}_2$ Butterfly Complexes

Jan Arneth,*[a]# Xian-Feng Li,[b]# Jonas Braun,*[b,c,d] Sagar Paul,[e] Michael Schulze,[e] Christopher E. Anson,[b] Wolfgang Wernsdorfer,[d,e] Annie K. Powell,*[b,c,d] Rüdiger Klingeler*[a]

# These authors have contributed equally

Corresponding authors: jan.arneth@kip.uni-heidelberg.de, jonas.braun2@kit.edu, annie.powell@kit.edu and ruediger.klingeler@kip.uni-heidelberg.de

**Table of Contents**

# 1. Experimental

## 1.1 Methods

Powder X-ray diffraction measurements were performed on a STOE STADI-P equipped with a Cu-Kα source with a characteristic wavelength of 1.5405 Å. Elemental analysis was performed on an Elementar Vario MicroCube. IR spectroscopy was performed on a Nicolet iS 50 with ATR attachment.

## 1.2 Synthesis

LnCl$_3$·6H$_2$O salts were produced using hydrochloric acid and the respective Ln oxide. All other reagents were obtained from commercial sources and used without further purification. All reactions were performed under the exclusion of oxygen using standard Schlenk techniques. Although the reaction solution is sensitive to oxidation, crystalline products were air stable.

**[V$_2$Ln$_2$(μ$_3$-OMe)$_2$(${}^t$Budea)$_2$(piv)$_6$]**

A solution of *N-tert*-butyl-diethanolamine (${}^t$Bu-H$_2$dea) (1 mmol, 161.3 mg) in 5 ml MeCN was added to a solution of VCl$_3$ (0.5 mmol, 78.7 mg), LnCl$_3$·6H$_2$O (0.5 mmol) and pivalic acid (Hpiv) (4 mmol, 296.5 mg) in 16 ml of a 1:1 mixture of MeOH and MeCN under Schlenk conditions. After 10 minutes of stirring, 1 ml of Et$_3$N was added and the resulting mixture stirred for 24 hours at RT. Subsequently, the solution was filtered and left undisturbed for crystallisation. The products were obtained as purple crystals suitable for single crystal X-ray diffraction.

**1$_Y$** was obtained in a yield of 96.5 mg (0.073 mmol = 29% based on Y$^{III}$).
**Anal. Calc (found) %** for V$_2$Y$_2$C$_{48}$H$_{94}$N$_2$O$_{18}$·2CH$_3$OH: C, 45.11 (44.87); H, 7.72 (7.55); N, 2.10 (2.01).
**Selected IR data (cm$^{-1}$)**: 2960 (m), 2926 (w), 2871 (w), 1560 (m), 1528 (w), 1481 (m), 1411 (m), 1375 (m), 1359 (m), 1226 (m), 1089 (w), 1089 (w), 1026 (w), 896 (w), 600 (w), 495 (w), 459 (w), 427 (w).

**2$_{Tb}$** was obtained in a yield of 198.6 mg (0.135 mmol = 54% based on Tb$^{III}$).
**Anal. Calc (found) %** for V$_2$Tb$_2$C$_{48}$H$_{94}$N$_2$O$_{18}$·2CH$_3$OH: C, 40.82 (40.73); H, 6.99 (7.03); N, 1.90 (1.87).
**Selected IR data (cm$^{-1}$)**: 2960 (m), 2925 (w), 1563 (m), 1524 (w), 1481 (m), 1458 (w), 1409 (m), 1375 (m), 1359 (m), 1260 (w), 1225 (m), 1086 (m), 1049 (w), 1018 (w), 894 (w), 793 (m), 598 (w), 494 (w), 457 (w), 422 (w).

**3$_{Dy}$** was obtained in a yield of 229.1 mg (0.155 mmol = 62% based on Dy$^{III}$).
**Anal. Calc (found) %** for V$_2$Dy$_2$C$_{48}$H$_{94}$N$_2$O$_{18}$·2CH$_3$OH: C, 40.63 (40.59); H, 6.95 (6.70); N, 1.90 (1.83).
**Selected IR data (cm$^{-1}$)**: 2960 (w), 2925 (w), 2870 (w), 1561 (m), 1527 (w), 1482 (m), 1409 (m), 1375 (w), 1359 (w), 1226 (w), 1088 (w), 1019 (w), 895 (w), 895 (w), 753 (w), 598 (w).

**4$_{Ho}$** was obtained in a yield of 167.6 mg (0.113 mmol = 45% based on Ho$^{III}$).
**Anal. Calc (found) %** for V$_2$Ho$_2$C$_{48}$H$_{94}$N$_2$O$_{18}$·2CH$_3$OH: C, 40.49 (40.38); H, 6.93 (6.87); N, 1.89 (1.82).
**Selected IR data (cm$^{-1}$)**: 2956 (m), 2926 (w), 2867 (w), 1591 (m), 1569 (s), 1533 (w), 1481 (m), 1428 (m), 1407 (s), 1374 (m), 1358 (m), 1225 (m), 1097 (m), 1082 (w), 1021 (m), 909 (w), 896 (w), 794 (w), 601 (w), 580 (w), 497 (w), 462 (w), 427 (w).

**5$_{Er}$** was obtained in a yield of 160.7 mg (0.108 mmol = 43% based on Er$^{III}$).
**Anal. Calc (found) %** for V$_2$Er$_2$C$_{48}$H$_{94}$N$_2$O$_{18}$·2CH$_3$OH: C, 40.37 (40.31); H, 6.91 (6.84); N, 1.88 (1.84).
**Selected IR data (cm$^{-1}$)**: 2962 (m), 2928 (w), 2871 (w), 1591 (m), 1571 (s), 1531 (w), 1481 (m), 1429 (m), 1409 (s), 1374 (m), 1359 (m), 1229 (m), 1099 (m), 1084 (w), 1027 (m), 909 (w), 899 (w), 798 (w), 601 (w), 585 (w), 497 (w), 466 (w), 427 (w).

**6<sub>Tm</sub>** was obtained in a yield of 183.4 mg (0.123 mmol = 49% based on Tm$^{III}$).
**Anal. Calc (found) %** for V$_2$Tm$_2$C$_{48}$H$_{94}$N$_2$O$_{18}$·2CH$_3$OH: C, 40.24 (40.15); H, 6.89 (6.74); N, 1.88 (1.74).
**Selected IR data (cm$^{-1}$)**: 2959 (w), 2926 (w), 1562 (m), 1530 (w), 1481 (m), 1410 (m), 1375 (w), 1359 (w), 1226 (w), 1090 (w), 896 (w), 601 (w).

**7<sub>Yb</sub>** was obtained in a yield of 119.9 mg (0.08 mg = 32% based on Yb$^{III}$).
**Anal. Calc (found) %** for V$_2$Yb$_2$C$_{48}$H$_{94}$N$_2$O$_{18}$·2CH$_3$OH: C, 40.05 (39.41); H, 6.85 (6.75); N, 1.87 (1.76).
**Selected IR data (cm$^{-1}$)**: 2961 (w), 2927 (w), 1563 (m), 1532 (w), 1482 (m), 1410 (m), 1375 (w), 1359 (w), 1225 (w), 1091 (w), 1020 (w), 603 (w).

Despite considerable efforts, it was unfortunately not possible to obtain the corresponding Gd$^{III}$-analogue in a pure form.

## 1.3 Crystallography

**Table S1.** Crystallographic data for compounds **1$_Y$** to **7$_{Yb}$**.

| Compound | 1$_Y$ | 2$_{Tb}$ | 3$_{Dy}$ | 4$_{Ho}$ |
|---|---|---|---|---|
| Formula | C$_{50}$H$_{102}$N$_2$O$_{20}$V$_2$Y$_2$ | C$_{50}$H$_{102}$N$_2$O$_{20}$Tb$_2$V$_2$ | C$_{50}$H$_{102}$Dy$_2$N$_2$O$_{20}$V$_2$ | C$_{50}$H$_{102}$Ho$_2$N$_2$O$_{20}$V$_2$ |
| FW / g mol$^{-1}$ | 1331.03 | 1471.05 | 1478.21 | 1483.07 |
| Crystal System | Triclinic | Triclinic | Triclinic | Triclinic |
| Space Group | $P\bar{1}$ | $P\bar{1}$ | $P\bar{1}$ | $P\bar{1}$ |
| $a$ / Å | 9.7025(3) | 9.7942(5) | 9.7345(3) | 9.8522(4) |
| $b$ / Å | 13.5515(4) | 13.5453(7) | 13.6421(4) | 13.4007(6) |
| $c$ / Å | 14.4081(5) | 14.4123(8) | 14.4772(4) | 14.3546(6) |
| α / ° | 115.388(3) | 115.277(4) | 115.179(2) | 115.041(3) |
| β / ° | 97.040(3) | 96.802(4) | 97.400(2) | 96.839(3) |
| γ / ° | 104.724(3) | 105.723(4) | 104.700(2) | 106.296(3) |
| $U$ / Å$^3$ | 1596.10(10) | 1602.91(16) | 1620.21(9) | 1585.22(12) |
| Z | 1 | 1 | 1 | 1 |
| T / K | 180(2) | 150(2) | 180(2) | 150(2) |
| $F$(000) | 700 | 752 | 754 | 756 |
| $D_c$ / Mg m$^{-3}$ | 1.385 | 1.524 | 1.515 | 1.554 |
| λ / Å | 1.54178 | 1.34143 | 0.71073 | 1.34143 |
| μ / mm$^{-1}$ | 5.300 | 13.293 | 2.625 | 13.056 |
| Data Measured | 13782 | 18150 | 21279 | 17314 |
| Unique Data | 5950 | 7556 | 10816 | 7483 |
| $R_{int}$ | 0.0221 | 0.0358 | 0.0838 | 0.0407 |
| Data with I ≥ 2σ(I) | 5816 | 7079 | 9654 | 7080 |
| $wR_2$ (all data) | 0.1104 | 0.1348 | 0.1459 | 0.1416 |
| S (all data) | 1.058 | 1.032 | 1.033 | 1.060 |
| $R_1$ [I ≥ 2σ(I)] | 0.0393 | 0.0497 | 0.0528 | 0.0508 |
| Param./Restr. | 425 / 85 | 397 / 35 | 403 / 42 | 395 / 43 |
| Biggest diff. peak and hole / eÅ$^{-3}$ | +1.01 / -1.20 | +2.04 / -2.83 | +3.49 / -3.02 | +2.67 / -2.13 |
| CCDC number | 2406392 | 2406393 | 2406394 | 2406395 |

**Table S1 (continued).** Crystallographic data for compounds **1$_Y$** to **7$_{Yb}$**.

| Compound | **5$_{Er}$** | **6$_{Tm}$** | **7$_{Yb}$** |
|---|---|---|---|
| Formula | $C_{50}H_{102}Er_2N_2O_{20}V_2$ | $C_{50}H_{102}N_2O_{20}Tm_2V_2$ | $C_{50}H_{102}N_2O_{20}V_2Yb_2$ |
| FW / g mol$^{-1}$ | 1487.73 | 1491.07 | 1499.29 |
| Crystal System | Triclinic | Triclinic | Triclinic |
| Space Group | $P\bar{1}$ | $P\bar{1}$ | $P\bar{1}$ |
| $a$ / Å | 9.7136(2) | 9.6975(3) | 9.7042(4) |
| $b$ / Å | 13.5673(3) | 13.5481(4) | 13.5421(7) |
| $c$ / Å | 14.3884(4) | 14.3920(4) | 14.3843(8) |
| α / ° | 115.124(2) | 115.060(2) | 115.042(4) |
| β / ° | 97.487(2) | 97.580(2) | 97.730(4) |
| γ / ° | 104.676(2) | 104.511(2) | 104.460(4) |
| $U$ / Å$^3$ | 1598.37(7) | 1595.96(9) | 1595.00(16) |
| $Z$ | 1 | 1 | 1 |
| T / K | 180(2) | 180(2) | 180(2) |
| $F$(000) | 758 | 760 | 762 |
| $D_c$ / Mg m$^{-3}$ | 1.546 | 1.551 | 1.561 |
| λ / Å | 1.34143 | 1.34143 | 1.34143 |
| μ / mm$^{-1}$ | 10.460 | 10.966 | 11.498 |
| Data Measured | 18607 | 18954 | 15961 |
| Unique Data | 7538 | 7489 | 6814 |
| $R_{int}$ | 0.0203 | 0.0179 | 0.0429 |
| Data with I ≥ 2σ(I) | 7043 | 6987 | 5872 |
| $wR_2$ (all data) | 0.0836 | 0.0645 | 0.1604 |
| $S$ (all data) | 1.042 | 1.029 | 1.106 |
| $R_1$ [I ≥ 2σ(I)] | 0.0305 | 0.0246 | 0.0560 |
| Param./Restr. | 437 / 160 | 436 / 124 | 414 / 37 |
| Biggest diff. peak and hole / eÅ$^{-3}$ | +0.68 / -0.68 | +0.66 / -0.93 | +1.86 / -1.96 |
| CCDC number | 2406396 | 2406397 | 2406398 |

**Table S2.** Selected bond lengths (Å) and angles (°) for **1$_Y$** to **7$_{Yb}$**.

|  | **1$_Y$** | **2$_{Tb}$** | **3$_{Dy}$** | **4$_{Ho}$** | **5$_{Er}$** | **6$_{Tm}$** | **7$_{Yb}$** |
|---|---|---|---|---|---|---|---|
| Ln1-O1 | 2.4345(16) | 2.461(3) | 2.461(2) | 2.449(3) | 2.4345(17) | 2.4250(15) | 2.418(4) |
| Ln1'-O1 | 2.4571(16) | 2.472(3) | 2.465(2) | 2.450(3) | 2.4476(17) | 2.4373(15) | 2.432(4) |
| Ln1-O2 | 2.2784(16) | 2.301(3) | 2.287(2) | 2.280(3) | 2.2620(17) | 2.2517(15) | 2.240(4) |
| Ln1-O3' | 2.2661(16) | 2.289(3) | 2.273(2) | 2.271(3) | 2.2556(17) | 2.2427(15) | 2.231(4) |
| Ln1-O5 | 2.3444(18) | 2.367(3) | 2.353(3) | 2.349(3) | 2.3318(19) | 2.3209(17) | 2.314(4) |
| Ln1-O7' | 2.3704(18) | 2.398(3) | 2.391(3) | 2.387(3) | 2.364(2) | 2.3535(18) | 2.349(4) |
| Ln1-O8 | 2.3680(19) | 2.412(4) | 2.390(3) | 2.384(4) | 2.373(2) | 2.3613(18) | 2.353(4) |
| Ln1-O9 | 2.3687(18) | 2.384(3) | 2.386(3) | 2.350(4) | 2.363(2) | 2.3566(17) | 2.343(4) |
|  |  |  |  |  |  |  |  |
| V1-O1 | 2.0516(17) | 2.057(3) | 2.050(2) | 2.052(3) | 2.0515(17) | 2.0480(15) | 2.049(4) |
| V1-O2 | 1.9266(16) | 1.931(3) | 1.933(2) | 1.931(3) | 1.9328(18) | 1.9298(16) | 1.929(4) |
| V1-O3 | 1.9262(16) | 1.931(3) | 1.936(2) | 1.923(3) | 1.9284(18) | 1.9268(16) | 1.925(4) |
| V1-O4 | 2.0258(17) | 2.018(3) | 2.024(2) | 2.016(3) | 2.0243(19) | 2.0235(17) | 2.031(4) |
| V1-O6 | 2.0111(17) | 2.018(3) | 2.014(3) | 2.022(3) | 2.0156(19) | 2.0153(17) | 2.008(4) |
| V1-N1 | 2.226(2) | 2.236(4) | 2.228(3) | 2.225(4) | 2.225(2) | 2.2231(19) | 2.220(5) |
|  |  |  |  |  |  |  |  |
| Ln1···Ln1' | 4.1285(4) | 4.1551(5) | 4.1496(3) | 4.1289(5) | 4.1187(3) | 4.1022(3) | 4.0916(6) |
| Ln1···V1 | 3.3326(5) | 3.3567(7) | 3.3427(5) | 3.3292(7) | 3.3164(4) | 3.3086(4) | 3.2963(9) |
| Ln1'···V1 | 3.3265(4) | 3.3556(7) | 3.3428(5) | 3.3272(8) | 3.3175(4) | 3.3080(4) | 3.2984(9) |
|  |  |  |  |  |  |  |  |
| Ln1-O1-Ln1' | 115.13(6) | 114.77(11) | 114.78(8) | 114.89(12) | 115.05(7) | 115.06(6) | 115.02(14) |
| V1-O1-Ln1 | 95.58(6) | 95.55(11) | 95.19(8) | 94.99(12) | 94.96(6) | 95.04(6) | 94.74(14) |
| V1-O1-Ln1' | 94.66(6) | 95.18(11) | 95.08(8) | 94.89(12) | 94.61(6) | 94.64(5) | 94.41(13) |
| V1-O2-Ln1 | 104.54(7) | 104.63(13) | 104.46(10) | 104.19(13) | 104.21(7) | 104.34(6) | 104.27(16) |
| V1-O3-Ln1' | 104.73(7) | 105.02(12) | 104.89(10) | 104.66(14) | 104.65(7) | 104.75(6) | 104.80(16) |

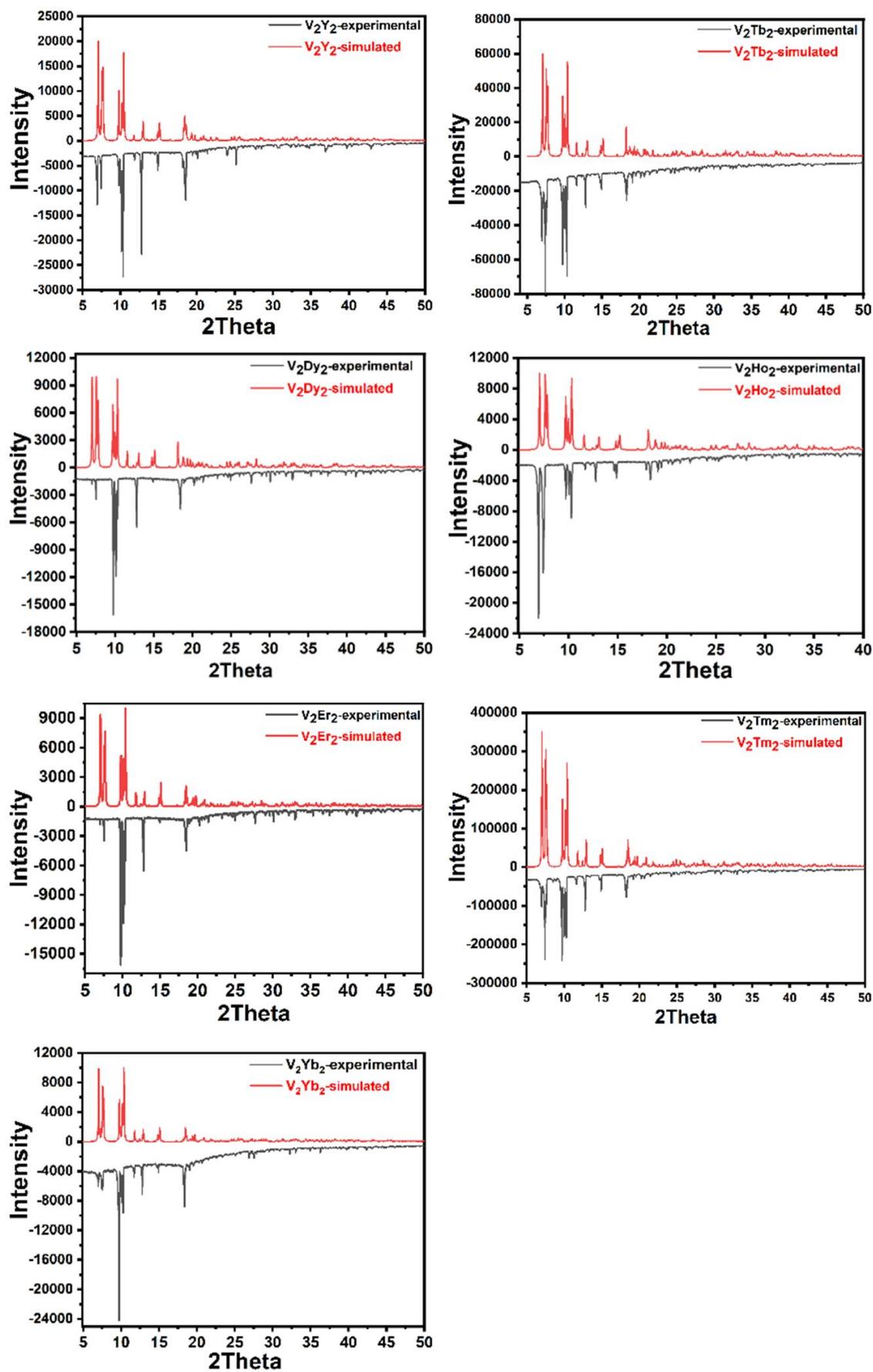

**Figure S1.** Experimental powder X-ray diffraction patterns and simulations using the single crystal structures and Mercury 2020.3.0.[63] for compounds **1_Y – 7_Yb**.

## 2. Bond Valence Sum (BVS) Analysis

**Table S3.** $V^{III}$ valences calculated from Bond Valence Sum Analysis[39] for V(1) in compounds **1$_Y$-7$_{Yb}$**.

| Compound | Calculated valence for V(1) |
|---|---|
| **1$_Y$** | 2.94 |
| **2$_{Tb}$** | 2.97 |
| **3$_{Dy}$** | 2.95 |
| **4$_{Ho}$** | 2.97 |
| **5$_{Er}$** | 2.96 |
| **6$_{Tm}$** | 2.97 |
| **7$_{Yb}$** | 2.97 |

## 3. Additional Magnetic Data

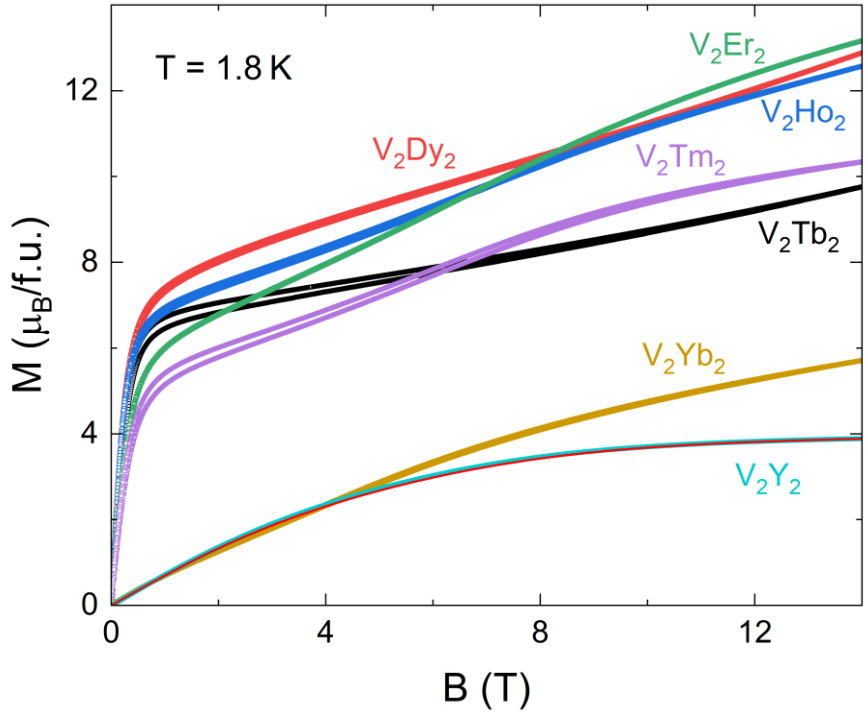

**Figure S2.** Isothermal magnetisation of $1_Y$ – $7_{Yb}$ at a constant temperature of T = 1.8 K. The red solid line depicts a simulation of $1_Y$ using the parameters obtained from HF-EPR (see the main text).

**Analysis of the magnetisation plateau in the microSQUID data:**

While the overall shape of the *M(B)* loops at 30 mK exhibits a predominantly ferromagnetic character, a magnetisation plateau appears near zero field as the bath temperature is increased (cf. inset of Fig. 7(b)). Such a temperature-induced transition indicates the presence of antiferromagnetic excited states several Kelvin above the ground state and can be rationalised with the help of the simulated Zeeman diagrams for $4_{Ho}$, $5_{Er}$ and $6_{Tm}$ presented in Figure S6. From the general trend it becomes evident that the antiferromagnetic state exhibits a small energy shift in zero field from an adjacent ferromagnetic state which leads to a level crossing at *B* < 0.5 T. As the spin temperature rises above 1 K, thermal population of the excited states via under-barrier temperature-dependent processes allows for the detection of the corresponding exchange field $B_{ex}$ ≈ 0.05(1) T. Such a small exchange field can likely be attributed to the weak dipolar Dy-Dy coupling $J_{Dy\text{-}Dy}$. Assuming an Ising-like character of the Dy moments, $J_{Dy\text{-}Dy}$ can be estimated using the exchange field approach $J_{Dy-Dy} = -\mu_B g_J \frac{B_{ex}}{m_J}$, with $g_J$ = 4/3 and $m_J$ = 15/2, yielding $J_{Dy\text{-}Dy}$ = -6(1) mK.[53]

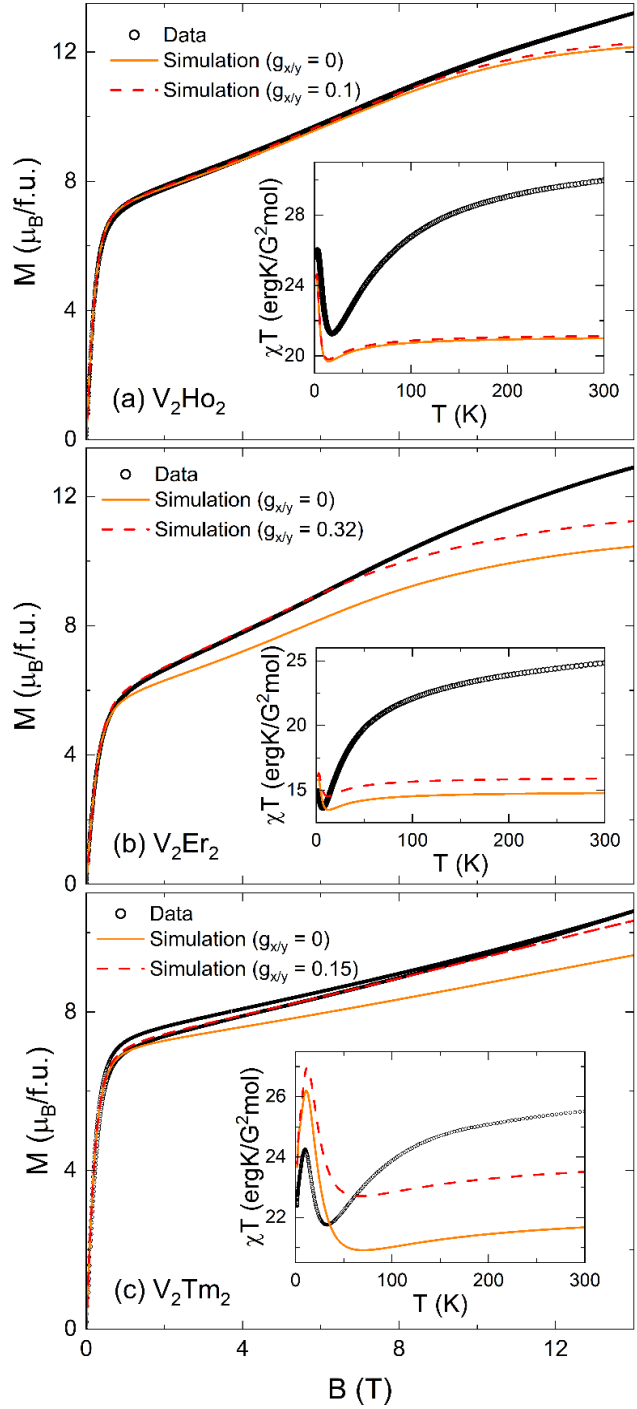

**Figure S3.** Isothermal magnetisation of **4_Ho** (a), **5_Er** (a) and **6_Tm** (c) at T = 1.8 K and simulations using the strict Ising model (orange solid line) in equation 2 and an almost-Ising model (red dashed line) with the parameters shown in Table 2 (see the main text).

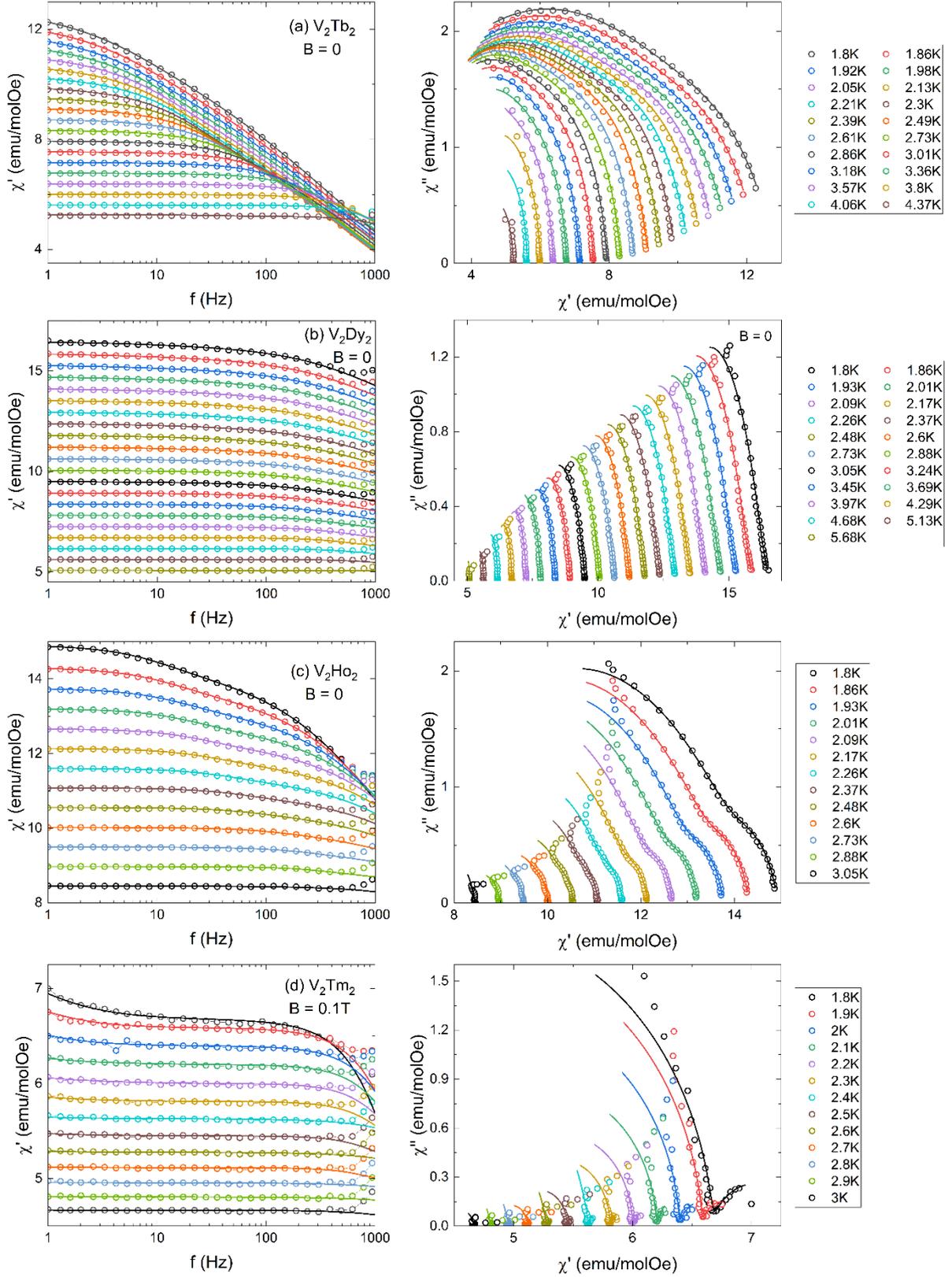

**Figure S4.** In-phase ac susceptibility (left) and Cole-Cole plots (right) of of **2**$_{Tb}$ (a) ,**3**$_{Dy}$ (b),**4**$_{Ho}$, (c) and **6**$_{Tm}$ (d) at selected temperatures. Solid lines depict fits of equation 3 to the data as described in the main text.

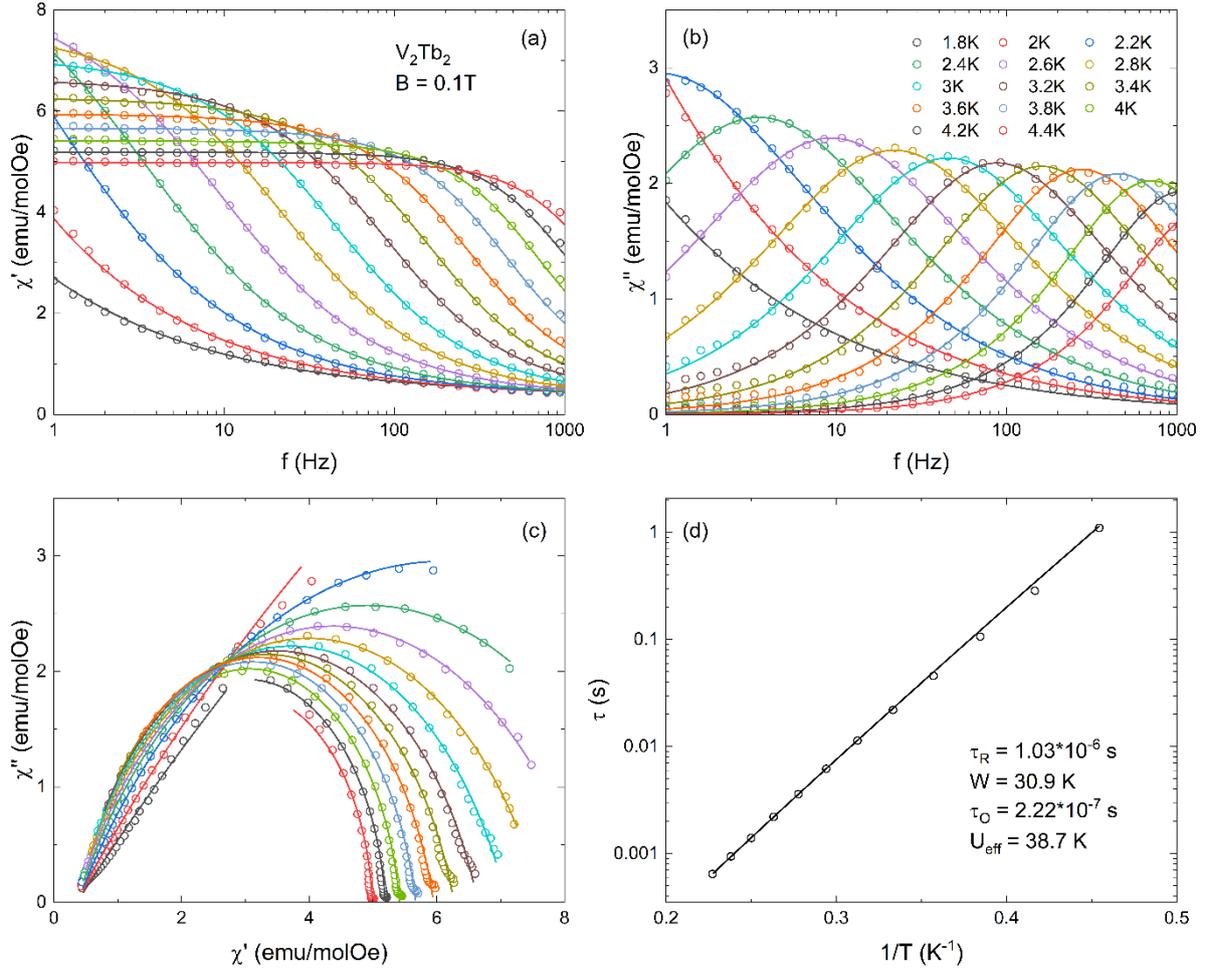

**Figure S5.** In-phase susceptibility (a), out-of-phase susceptibility (b), Cole-Cole plot (c) and Arrhenius plot of the obtained relaxation times at $H_{ext}$ = 0.1 T of **2$_{Tb}$**. Solid lines in (a), (b) and (c) depict fits of equation S1 to the data. The solid line in (d) depicts a fit of equation 4 (see the main text) with the parameters shown in the plot.

$$\chi_{ac}(f) = \left(\chi_{S,n} + \frac{\chi_T - \chi_S}{1+(if\tau)^{1-\alpha}}\right) \qquad (S1)$$

## 4. Additional HF-EPR Data

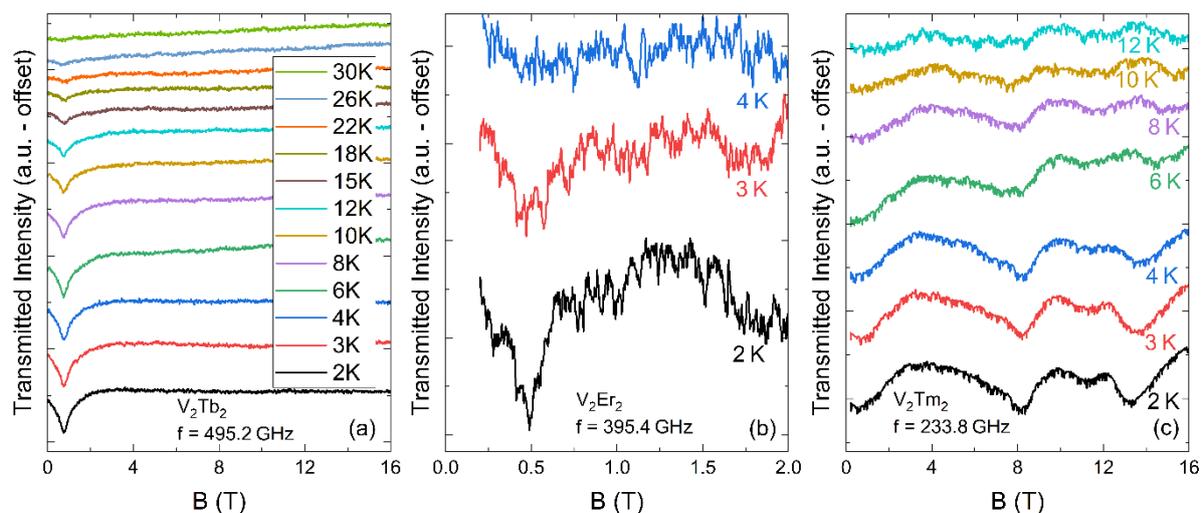

**Figure S6.** Loose powder HF-EPR spectra of **2**$_{Tb}$ (a), **5**$_{Er}$ (b) and **6**$_{Tm}$ (c) at fixed frequencies and variable temperatures.

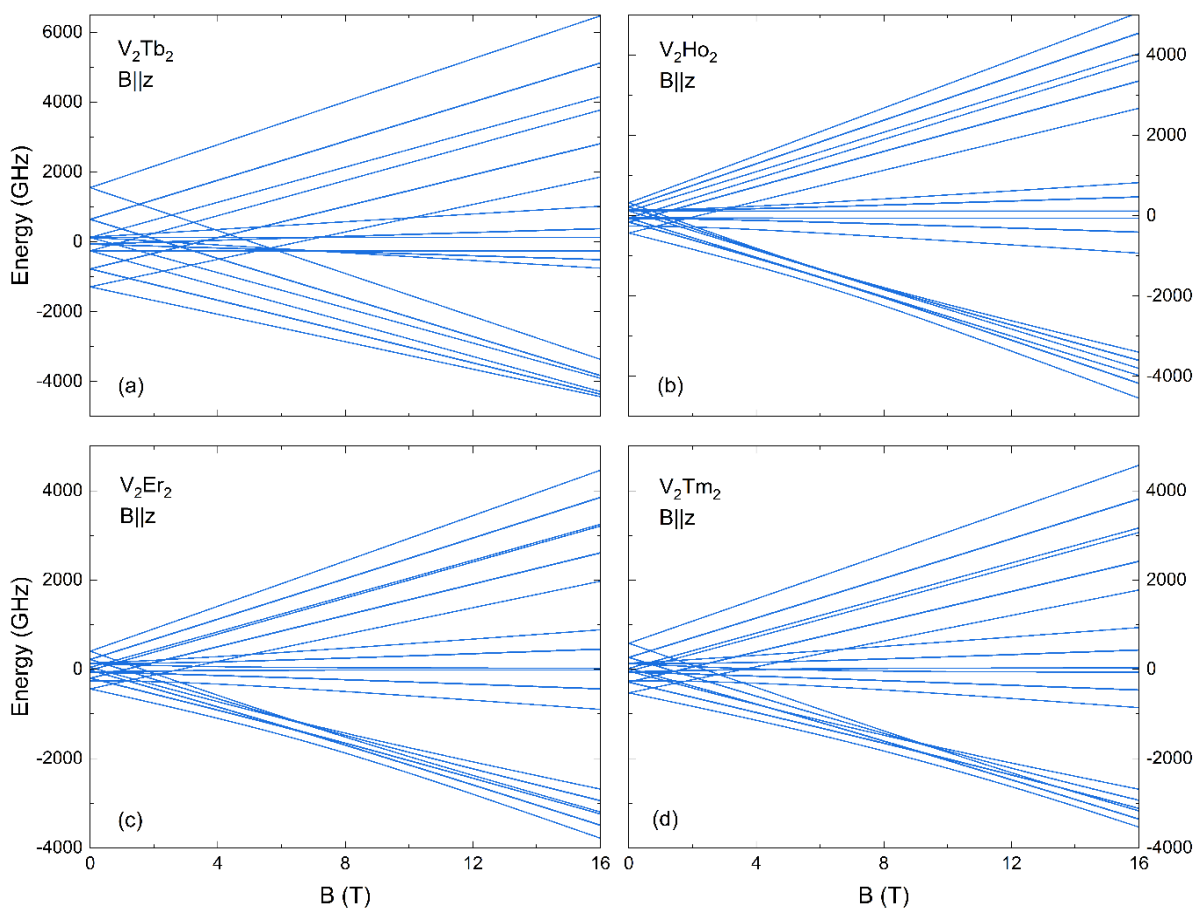

**Figure S7.** Simulated energy level diagram of **2**$_{Tb}$ (a), **4**$_{Ho}$ (b), **5**$_{Er}$ (c) and **6**$_{Tm}$ (d) for B||z using the spin Hamiltonian parameters listed in Tab. 2.

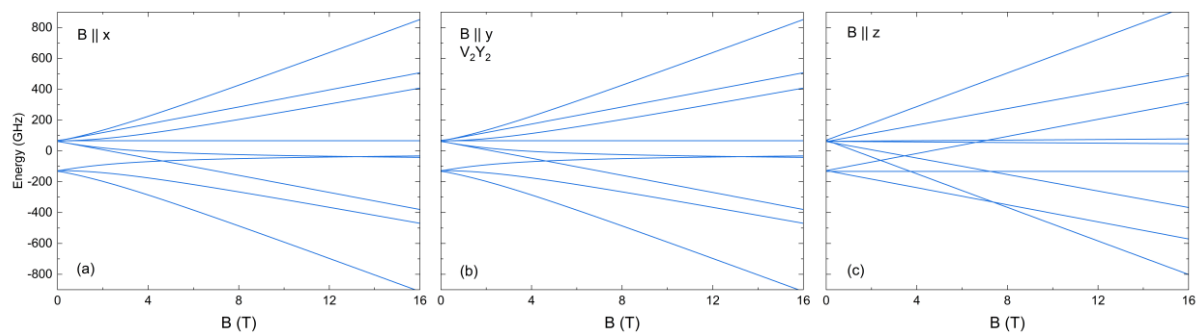

**Figure S8.** Simulated energy level diagram of **1**$_Y$ for B||x (a), B||y (b) and B||z (c) using the spin Hamiltonian parameters as described in the main text.